\documentclass[prd,onecolumn,11pt,floatfix,altaffilletter,superscriptaddress,preprintnumbers,tightenlines,showpacs,showkeys,nofootinbib]{revtex4-2}
\pdfoutput=1
\usepackage[utf8]{inputenc}
\usepackage{graphicx,xcolor}
\usepackage{hyperref}
\usepackage[T1]{fontenc}
\usepackage{multirow}
\usepackage{mathtools}
\usepackage{csquotes}
\usepackage[english]{babel}
\usepackage{subcaption}
\usepackage{amsmath}
\usepackage{amssymb}
\usepackage{siunitx}
\usepackage{xspace}
\usepackage{enumitem}
\usepackage[capitalise]{cleveref}

\usepackage{xpatch}
\makeatletter
\patchcmd{\@ssect@ltx}
    {\addcontentsline{toc}{#1}{\protect\numberline{}#8}}
    {}
    {}
    {}
\makeatother

\makeatletter
\renewcommand\@makecaption[2]{%
  \par
  \vskip\abovecaptionskip
  \begingroup
   \small\rmfamily
    \begingroup
     \samepage
     \flushing
     \let\footnote\@footnotemark@gobble
     \@make@capt@title{#1}{#2}\par
    \endgroup
  \endgroup
  \vskip\belowcaptionskip
}
\makeatother

\DeclareSIUnit\year{yr}
\DeclareSIUnit\years{yrs}

\newcommand{\ev}[1]{\ensuremath{\left\langle #1 %
                     \right\rangle}} 

\DeclareMathOperator{\arctanh}{arctanh}

\newcommand{\diff}{\textrm{d}}
\newcommand{\tbh}{\ensuremath T_\text{BH}\xspace}
\newcommand{\mbh}{\ensuremath M_\text{BH}\xspace}

\newcommand{\timebh}{\ensuremath t_\text{BH}\xspace}

\newcommand{\mpl}{\ensuremath M_\text{Pl}\xspace}

\renewcommand{\vec}[1]{{\mathbf{#1}}}
\newcommand{\iso}[2]{{\ensuremath{{}^{#2}}\ensuremath{\rm #1}}}

\begin{document}

\title{Dark Matter, Destroyer of Worlds: Neutrino, Thermal, and Existential Signatures from Black Holes in the Sun and Earth}

\author{Javier F. Acevedo}
\email{javier.fernandezacevedo@queensu.ca}
\affiliation{The McDonald Institute and Department of Physics, 
            Queen's University, Kingston, Ontario, Canada}
             
\author{Joseph Bramante}
\email{joseph.bramante@queensu.ca}
\affiliation{The McDonald Institute and Department of Physics, 
            Queen's University, Kingston, Ontario, Canada}
\affiliation{Perimeter Institute for Theoretical Physics, 
            Waterloo, Ontario, Canada}

\author{Alan~Goodman}
\email{alan.goodman@queensu.ca}
\affiliation{The McDonald Institute and Department of Physics, 
            Queen's University, Kingston, Ontario, Canada}
            
\author{Joachim Kopp}
\email{jkopp@cern.ch}
\affiliation{Theoretical Physics Department, CERN,  
1211 Geneva, Switzerland}
\affiliation{PRISMA Cluster of Excellence \& Mainz Institute for
             Theoretical Physics, \\
             Johannes Gutenberg University,
             55099 Mainz, Germany}

\author{Toby Opferkuch}
\email{toby.opferkuch@cern.ch}
\affiliation{Theoretical Physics Department, CERN,  
1211 Geneva, Switzerland}

\date{\today}
\preprint{CERN-TH-2020-209, MITP/20-081}

\begin{abstract}
Dark matter can be captured by celestial objects and accumulate at their centers, forming a core of dark matter that can collapse to a small black hole, provided that the annihilation rate is small or zero. If the nascent black hole is big enough, it will grow to consume the star or planet. We calculate the rate of dark matter accumulation in the Sun and Earth, and use their continued existence to place novel constraints on high mass asymmetric dark matter interactions. We also identify and detail less destructive signatures: a newly-formed black hole can be small enough to evaporate via Hawking radiation, resulting in an anomalous heat flow emanating from Earth, or in a flux of high-energy neutrinos from the Sun observable at IceCube. The latter signature is entirely new, and we find that it may cover large regions of parameter space that are not probed by any other method.
\end{abstract}

\maketitle

\vspace{-0.5cm}
\renewcommand{\baselinestretch}{0.85}\normalsize
\tableofcontents
\renewcommand{\baselinestretch}{1.0}\normalsize

\section{Introduction}
\label{sec:intro}

The existence of dark matter (DM) has been established through its gravitational interactions with visible matter, observed in the way galaxies rotate, through lensing of light in galaxy clusters, and by its imprint on the structure of the early universe.  However, the mass of dark matter and its hypothesized non-gravitational interactions remain a compelling mystery for physics and astronomy. One approach towards resolving this mystery is to search for the potential impact of dark matter on astrophysical objects. In this work, we study dark matter gravitationally captured by the Sun and the Earth, focusing on the conditions under which dark matter can form black holes in their interiors. These black holes could be detected through their Hawking radiation, or through the destruction of the host body. As, evidently, neither the Sun nor the Earth has suffered this fate yet, we will be able to set limits on dark matter properties.

Dark matter particles passing through the Sun or Earth can lose some of their kinetic energy by scattering off ordinary matter. Particles that lose enough energy during their transit are nudged into a closed gravitational orbit that intersects the star or planet. Through subsequent and repeated scattering, the captured dark matter population can attain a temperature in equilibrium with the surrounding ordinary matter, forming a ``thermalized'' dark matter core deep inside the Sun or Earth. As more dark matter accumulates, this dark matter core can grow so massive that it collapses under its own gravity and forms a black hole, provided that repulsive forces such as thermal pressure and degeneracy pressure are overcome. The black hole can then either grow by accretion of surrounding material, or evaporate in the form of Hawking radiation. The formation of a black hole primarily depends on three relevant physical conditions:
\begin{enumerate}
  \item The dark matter core must have reached a critical mass to become self-gravitating and collapse under its own weight, thus overcoming both the thermal pressure generated by the surrounding ordinary matter, as well as quantum degeneracy pressure. To surpass quantum degeneracy pressure, enough dark matter mass must be accumulated such that it cannot be stabilized by this degeneracy pressure, much like the Chandresekhar threshold for black hole formation in neutron stars and white dwarfs. This dark Chandresekhar threshold will apply both to fermionic dark matter, as well as bosonic dark matter with an $\mathcal{O}(1)$ quartic self-coupling.
  
  Note that the requirement of reaching a critical total dark matter mass inside the star immediately implies that the dark matter candidates to which our bounds apply must be non-annihilating or very weakly annihilating.  If the dark matter had a sizeable annihilation cross-section, it would be depleted by annihilation long before reaching the critical mass. The most natural scenarios in which dark matter annihilation is avoided are so-called ``asymmetric dark matter'' models \cite{Kaplan:2009ag, Petraki:2013wwa, Zurek:2013wia}, in which the dark matter sector features a particle--anti-particle asymmetry similar to the baryon asymmetry in the Standard Model (SM) sector. This asymmetry implies that only dark particles are abundant in the Universe today, while dark anti-particles have annihilated away shortly after the Big Bang.
    
  \item In order for collapse to occur, the captured dark matter must have lost enough energy via scattering to reach thermal equilibrium over timescales shorter than the age of the solar system. Therefore, we require the scattering cross-section of dark matter on ordinary matter to be sufficiently large such that this thermalization process occurs quickly. At the same time, there is an upper limit to the cross-section since a very large dark matter cross-section will cause the dark matter to be trapped in the outer layers of the Sun or the Earth.
    
  \item After it begins collapsing, the gravitationally unstable dark matter core must be able to shed its gravitational potential energy in order to continue collapsing. Here we will simply consider dark matter scattering with nuclei as a guaranteed and minimal process to achieve this.
\end{enumerate}
After a black hole forms from accumulated dark matter, it will either evaporate or destroy its host. Less massive dark matter tends to form the bigger, host-consuming black holes because lower-mass particles are more strongly affected by thermal and quantum pressure, so that more dark matter needs to be accumulated before its gravitational potential is able to overcome pressure forces to form a black hole. Such large back holes will grow rapidly by accretion of host material and newly-captured dark matter. Therefore, for lower dark matter masses, we can set bounds on the DM--nucleus scattering cross-section by requiring that the formation of a destructive black hole takes longer than the age of the solar system, approximately $5$~billion years. 

Heavier dark matter, on the other hand, will produce light black holes that evaporate quickly via Hawking radiation. This radiation can in principle be detected, either directly or by the energy it deposits in the surrounding matter.  The only component of Hawking radiation that may be directly detectable are neutrinos. All other particle species get absorbed before they are able to leave the dense environment in which the black hole was formed. As they are absorbed, however, they heat up the surrounding matter, and this extra heat may be detectable as it is radiated off into space. Obviously, this is only possible in planets whose heat flux from standard sources is relatively low. Because Earth's heat emanations and interior composition are well studied, Earth's heat flow will place leading bounds on some high mass dark matter that forms evaporating black holes. 

Dark matter that collapses to form black holes in astrophysical hosts has been studied in some detail for the cases of neutron stars and white dwarfs, \cite{Goldman:1989nd, Gould:1989gw, Kouvaris:2007ay, Bertone:2007ae, deLavallaz:2010wp, Kouvaris:2010vv, McDermott:2011jp, Kouvaris:2011fi, Kouvaris:2011gb, Bramante:2013hn, Bell:2013xk, Bramante:2014zca, Bramante:2015cua, Bramante:2016mzo, Bramante:2017ulk, Garani:2018kkd, Kouvaris:2018wnh, Kopp:2018jom, Acevedo:2019gre, Janish:2019nkk, East:2019dxt,Tsai:2020hpi,Takhistov:2020vxs,Dasgupta:2020mqg} (for related thermal signatures of dark matter in stars, see $e.g.$ \cite{McCullough:2010ai,Baryakhtar:2017dbj,Raj:2017wrv,Bell:2018pkk,Hamaguchi:2019oev,Camargo:2019wou,Bell:2019pyc,Garani:2019fpa,Joglekar:2019vzy,Acevedo:2019agu,Dasgupta:2019juq,Joglekar:2020liw,Garani:2020wge,Bell:2020lmm,Dasgupta:2020dik}). Previously, the possibility that heavy dark matter could form a black hole in the Earth was mentioned in Ref.~\cite{Starkman:1990nj}, where the authors noted a dark matter mass threshold beyond which black holes might form. Here, we treat the formation and evolution of black holes in the Sun and Earth for the first time and use these calculations to obtain new bounds on dark matter. In particular, we derive limits on dark matter interactions with nuclei based on the continuing existence of the Sun and Earth, on the non-observation of excessive heat flow from the Earth, and on the non-observation of the neutrino component of Hawking radiation from evaporating black holes. The new Earth heat flow bounds we obtain on non-annihilating {\em asymmetric} dark matter that forms evaporating black holes nicely complement prior Earth heat flow bounds on {\em symmetric} dark matter that self-annihilates \cite{Mack:2007xj, Bramante:2019fhi}.

The outline of the paper is as follows: we begin in \cref{sec:capture} with the computation of the dark matter capture rates in the Sun and the Earth, followed in \cref{sec:thermalization} by a discussion of the dark matter thermalization process. In \cref{sec:collapse}, we then derive the conditions under which the dark matter core collapses, and the evolution of the resulting black hole. In \cref{sec:neutrino}, we elaborate on neutrino signatures of evaporating black holes, before presenting and discussing our results in \cref{sec:results}. We summarize and conclude in \cref{sec:conclusions}.

\section{Dark Matter Capture}
\label{sec:capture}

As a dark matter particle traverses a star or planet, it can scatter against its constituent elements, losing a fraction of its kinetic energy with each successive scatter. If the dark matter particle's velocity drops below the escape velocity of the stellar body, then it will become gravitationally bound (or be ``captured'').  This is equivalent to saying that capture occurs when the energy loss due to scattering is larger than the kinetic energy which the dark matter particle had when it was far away from the object, before it fell into its gravitational well.

\subsection{Dark matter models and energy loss in scattering}
\label{sec:models}

Before proceeding, it will be helpful to outline the classes of dark matter models considered in this paper. First, as already mentioned in the introduction, dark matter that can collect in large enough quantities to form black holes in stars and planets tends to be non-annihilating or asymmetric \cite{Zurek:2013wia, Petraki:2013wwa,Kaplan:2001qk}. For other types of dark matter, depletion through self-annihilation is often efficient enough to prevent black hole formation. Therefore, for simplicity, we will assume hereafter that dark matter is asymmetric.

We will consider two types of dark matter--nucleon couplings. The first are spin-independent couplings \cite{LEWIN199687}, for which the dark matter scattering cross-section $\sigma_{\chi j}$ on a nucleus of mass number $A_j$ is related to the DM--nucleon cross-section $\sigma_{\chi N}$ through the relation
\begin{align}
    \sigma_{\chi j} = A_j^2 \bigg( \frac{\mu(m_j)}{\mu(m_N)} \bigg)^2 \sigma_{\chi N}
                    \xrightarrow{m_\chi \gg m_j} A_j^4 \sigma_{\chi N} \,.
    \label{eq:crossec}
\end{align}
Here, $\mu(m_N)$ ($\mu(m_j)$) is the reduced mass of the DM--nucleon (DM--nucleus) system, and the index $j$ labels the different nuclear scattering targets. In the rightmost expression, we have used the fact that $m_j \ll m_\chi$ for the dark matter masses of interest to us. We do not need to explicitly include nuclear form factors here as the typical momentum exchange during dark matter capture is much smaller than the inverse of a nuclear radius. The salient feature of \cref{eq:crossec} for our subsequent discussion is the scaling of the DM--nucleus cross-section with $A_j^4$. Because of this scaling with the nuclear mass we will in the following refer to this class of models as ``isotope-dependent''. Dark matter moving at velocity $v$ through a medium in which the number density of the $j$-th species of nuclei is $n_j$ will lose energy at a rate given by 
\begin{align}
    \frac{dE}{dt} &\simeq \sum_j n_j \sigma_{\chi j } v \, \bigg( \frac{[\mu(m_j)]^2}{m_j} v^2 \bigg) \,.
        & \text{(isotope-dependent DM--nucleus scattering)}
    \label{eq:losspernuc}
\end{align}
The term in parentheses gives the average energy loss the dark matter particle suffers in a single DM--nucleus scattering process \cite{Gould:1991hx,Acevedo:2019gre}.

Besides an isotope-dependent DM--nucleus coupling, we also consider the possibility that the dark matter scattering cross-section is independent of the nuclear properties. An important example for this scenario is dark matter in the form of large composite states that scatter elastically with all nuclei they encounters. The formation and subsequent cosmological evolution of large asymmetric composite dark matter states has received a great deal of attention in the recent literature, see for instance Refs.~\cite{Wise:2014ola, Hardy:2014mqa, Gresham:2017cvl}.  If the composite dark matter particles are physically much larger than the nuclei on which they scatter, and assuming all scattering is elastic, the contact scattering cross-section $\sigma_c$ is the same for all nuclei. We will refer to this type of dark matter as ``isotope-independent''. This contact interaction for large composite dark matter was also explored in, e.g., Refs.~\cite{Starkman:1990nj, Jacobs:2014yca, Cappiello:2020lbk,Bhoonah:2020dzs}. The rate of kinetic energy loss experienced by a dark matter particle with isotope-independent scattering cross-section moving through nuclei with number density $n_j$ will once again be given by \cref{eq:losspernuc}. In the limit $m_\chi \gg m_j$, this expression can now be further simplified to
\begin{align}
    \frac{dE}{dt} &\xrightarrow{m_\chi \gg m_j} \rho_* \sigma_{c } v^3 \,,
        & \text{(isotope-independent DM--nucleus scattering)}
    \label{eq:losscomp}
\end{align}
a result which highlights the fact that the energy loss in isotope-independent DM--nucleus scattering will depend only on the density of baryonic matter, $\rho_*$, but not on its chemical composition.

\subsection{Density, chemical composition, and temperature profiles of the Sun and Earth}
\label{sec:composition}

To accurately compute the dark matter energy loss rate via nuclear scattering, which determines its capture and thermalization, we need as inputs the density, chemical composition, and temperature profiles of the Sun and Earth. The density and composition determine the number density of scattering targets and the reduced mass of the DM--nucleus system, while the temperature profile controls the relative velocity of the DM--nucleus system and also the final dark matter distribution after thermalization is complete. We can neglect atomic thermal velocities as long as the dark matter is moving considerably faster than any terrestrial or solar constituent, but not when the dark matter and atomic velocities become comparable.

\begin{figure}
    \centering
    \includegraphics[width=\linewidth]{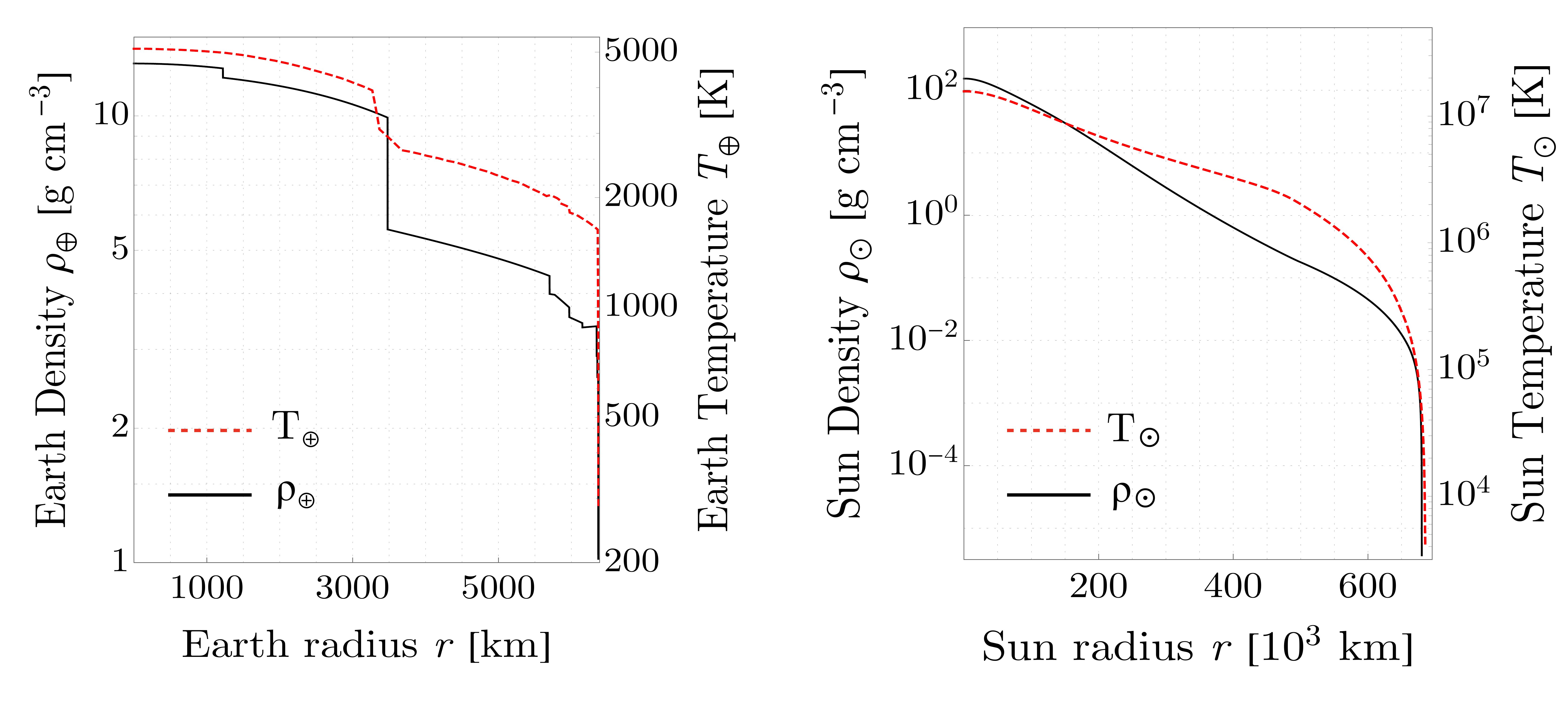}
    \caption{The solar and terrestrial density (black) and temperature (blue) profiles used in this work. These are taken from Refs.~\cite{Bahcall:2004pz} for the Sun, and from Refs.~\cite{Dziewonski:1981xy,JAUPART2015223, arevalo:2009, earle2015} for the Earth. The Sun and Earth radii are $R_{\odot} \approx \SI{6.96e10}{cm}$ and $R_{\oplus} \approx \SI{6.371e8}{cm}$, respectively.}
    \label{fig:DensityTemperature}
\end{figure}

We use the solar density and temperature profiles from Ref.~\cite{Bahcall:2004pz}, and take the Sun as made of 27.8\% $^4$He by number, with the rest being $^1$H \cite{Serenelli:2010fk}. Using this $^4$He fraction, which corresponds to the initial helium abundance when the solar system was formed, is justified because we are considering dark matter capture starting at the earliest stages of stellar evolution. For the Earth, we use the same model as in Ref.~\cite{Bramante:2019fhi}, based on the density profile of the Preliminary Reference Earth Model (PREM) from Ref. \cite{Dziewonski:1981xy}, the compositional profiles of Refs. \cite{clarke1924composition,WANG2018460,Morgan6973,MCDONOUGH2003547,johnston1974}, and a temperature profile comprised of the maximum temperature of Refs.~\cite{JAUPART2015223, arevalo:2009, earle2015} for any given radius. We use the maximum temperature because this will yield conservative bounds. More precisely, for higher temperatures dark matter will thermalize at larger radii, meaning more dark matter will need to be captured to form a black hole. Finally, for simplicity, we assume both the Sun and Earth to be perfect spheres with isotropic distributions of density, chemical composition, and temperature. The solar and terrestrial density and temperature profiles that enter our calculations are shown in \cref{fig:DensityTemperature}.

\subsection{Calculation of the capture rate}
\label{sec:capture-rate}

The capture of dark matter particles in stars and planets has been an active research topic since the 1980s \cite{Silk:1985ax,Press:1985ug, Krauss:1985ks, Srednicki:1986vj}, with a number of papers developing a detailed formalism for calculating the capture rate, including both capture through single scattering and through multiple scattering \cite{Gould:1987ir,Gould:1987ju,  Gould:1991va}. Here, we follow a slightly more modern semi-analytic approach (outlined in Ref.~\cite{Bramante:2019fhi}), which allows in particular for multiple scattering along different trajectories, with variable material composition along the trajectory. In this approach, we assume the following:
\begin{enumerate}
  \item Gravitational effects on dark matter trajectories are negligible prior to capture. This is a conservative assumption because gravity will pull dark matter particles closer towards the center of the star or planet, increasing its path length inside and causing it to traverse a denser medium. This phenomenon, sometimes called dark matter focusing, is discussed for instance in Refs.~\cite{Gould:1987ir, Goldman:1989nd,Peter:2009mi,Lee:2013wza,Bramante:2017xlb}.

  \item Scattering of dark matter on nuclei does not affect its trajectory. This means that, if the dark matter particle scatters multiple times on its way through the star or planet, these scatters occurs along a straight line. This is also a conservative assumption because a trajectory that more closely resembles a random walk will almost certainly result in a longer path length in matter than a straight trajectory, increasing the capture rate. The straight line approximation is also justified by the fact that we consider very heavy dark matter, $m_\chi \gg m_j$, implying that the momentum transfer in each individual scattering process is small.
\end{enumerate}
Far away from the stellar body, we assume the dark matter particles to have a flux-normalized distribution \cite{LEWIN199687}
\begin{align}
    f_\infty(\tilde{v}) = \frac{2 \tilde{v}^3}{v_0^4} \exp\bigg(- \frac{\tilde{v}^2}{v_0^2} \bigg)\,,
    \label{eq:boltzmann}
\end{align}
corresponding to a flux-weighted Maxwell--Boltzmann distribution. Here, $\tilde{v}$ denotes the velocity in the rest-frame of the galaxy, and $v_0 = \SI{220}{km/s}$ characterizes the dark matter velocity dispersion \cite{Bovy_2012,Read:2014qva,Pato:2015dua}. We have set the normalization of $f_\infty(\tilde{v})$ such that $\int_0^\infty f_\infty(\tilde{v}) \, \diff\tilde{v} = 1$. For our purpose of calculating heavy dark matter capture on the Earth and Sun, we must make a few improvements to \cref{eq:boltzmann}. First, we use a Heaviside $\Theta$ function of the form $\Theta(v_{eg} - \tilde v)$ to ensure that the dark matter velocity is below the galactic escape velocity, $v_{eg} = \SI{528}{km/s}$ at the Sun's location, $\sim \SI{8.3}{kpc}$ away from the Galactic Center~\cite{Deason_2019}. Second, to account for the velocity $\vec v_\text{rf}$ of the solar system with respect to the rest-frame of the galaxy, we need to shift the Boltzmann exponent in the standard way \cite{LEWIN199687} by replacing $\tilde{v} \to |\vec{v} + \vec{v}_\text{rf}|$. Here, $\vec{v}$ is the dark matter velocity in the Earth rest frame, and as usual we use bold face to denote 3-vectors and italics for their moduli, e.g.\ $v = |\vec{v}|$. There is also a corresponding shift for the $\Theta$ function to enforce the galactic escape velocity cutoff \cite{Bramante:2016rdh}. Finally, as dark matter particles approach a star or planet, they are accelerated in its gravitational field so that, upon reaching the surface, they are moving \textit{at least} at the escape velocity, $v_e \approx \SI{42}{km/s}$ for the Sun and $v_e \approx \SI{11.2}{km/s}$ for the Earth \cite{Peter:2009mi}. To account for this, we must shift the modulus of the dark matter velocity according to $v^2 \to v^2 - v_e^2$, and use a Heaviside $\Theta$ function to exclude speeds less than $v_e$. Consequently, the flux-weighted velocity distribution at the surface, now expressed in the Sun's or Earth's rest frame, is
\begin{align}
    f_*(\vec{v}) = \frac{(v^2-v_e^2)^{3/2}}{N_*} \exp\bigg(- \frac{\tilde{v}^2}{v_0^2} \bigg) \,
                   \Theta(v-v_e) \, \Theta(v_{eg}-\tilde{v}) \,,
    \label{eq:boltzmannShift}
\end{align}
with
\begin{align}
    \tilde{v}^2 \equiv v^2 - v_e^2 + v_{rf}^2 + 2 v_{rf} \sqrt{v^2-v_e^2} \, \cos\phi
    \label{eq:v-transform}
\end{align}
denoting the initial velocity (in the galactic rest frame) of a dark matter particle that reaches the surface of the Sun or Earth with a velocity $v$ (in the Sun's or Earth's rest frame). In \cref{eq:boltzmannShift}, $N_*$ is a normalizing factor that enforces $\int_0^\infty f_*(\vec{v}) \, \diff v\,\diff\cos{\phi} = 1$, and $\phi$ is the angle between the dark matter velocity vector $\vec{v}$ and the velocity of the solar system relative to the galactic rest frame, where we take $v_\text{rf} = |\vec{v}_\text{rf}| = \SI{230}{km/s}$ \cite{Bovy_2012,Read:2014qva}. Using the shifted distribution from \cref{eq:boltzmannShift} instead of the unshifted Maxwell--Boltzmann distribution is a departure from some analyses in the literature (see for instance Ref. \cite{Albuquerque:2000rk}). The local gravitational acceleration of the dark matter particles to a velocity of at least $v_e$ has a dramatic effect on the dark matter capture rate when only a small fraction of dark matter (namely the particles with the lowest velocities) are efficiently captured. If capture is efficient for all dark matter particles, the difference between $v^2$ and $v^2-v_e^2$ is on average negligible, see \cref{app:cap} for a more detailed discussion.

\begin{figure}
    \centering
    \includegraphics[width=0.3\linewidth]{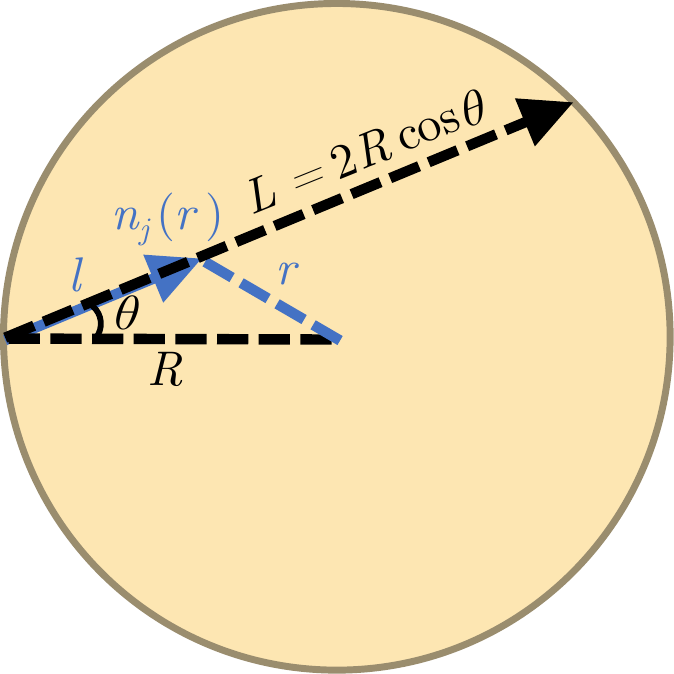}
    \caption{A schematic representation of the trajectory of a dark matter particle entering the Sun or Earth with an angle $\theta$ as measured from the normal to the surface.}
    \label{fig:SunSchematic}
\end{figure}

The distance $L$ a dark matter particle travels through the Sun or Earth is determined by its angle of entry $\theta$, defined as the angle between the dark matter's velocity and the vector normal to the surface. As can be easily deduced from the sketch in \cref{fig:SunSchematic}, the relation between $L$ and $\theta$ is $L = 2 R_* \cos\theta$, where $R_*$ is the radius of the star or planet. When travelling through a material of constant number density, $n_j$, a dark matter particle will scatter an average of 
\begin{align}
    \tau_j = n_j \sigma_{\chi j} L
    \label{eq:taujconst}
\end{align}
times against an element of the species indexed by $j$. $\tau_j$ is called the optical depth of the particle through the medium. In the case of a radially symmetric density profile, $n_j \equiv n_j(r)$, this equation becomes
\begin{align}
    \tau_j = \int_0^{L} n_j(r)\sigma_{\chi j} \, \diff l \,,
    \label{eq:tauj}
\end{align}
where $r = \sqrt{l^2 + R_*^2 - 2 l R_*\cos\theta}$ is the distance from the star's or planet's center at a given point $l$ along the trajectory.  

After a single scatter, the final state velocity $v_f$ of the dark matter particle as a function of its initial velocity $v_i$ is
\begin{align}
  v_f^\text{single} = v_i \sqrt{1 - z \beta_j^+},
  \label{eq:singleScatter}
\end{align}
where 
\begin{align}
    \beta^+_j \equiv \frac{4 m_j m_\chi}{(m_j + m_\chi)^2} \,,
    \label{eq:betapm}
\end{align}
and $z \in [0,1]$ is a kinematic factor determined by the scattering angle. We set $z \equiv 1/2$ in the following~\cite{Bramante:2018qbc, Bramante:2019fhi}. Combining \cref{eq:tauj,eq:singleScatter}, we find that the final velocity of a dark matter particle after traversing the whole star or planet is
\begin{align}
  v_f = v_i \prod_j (1-z\beta^+_j)^{\tau_j/2} \,.
  \label{eq:multiscatterVf}
\end{align}
For the particle to be captured, we require $v_f < v_e$. The maximum initial velocity (upon entering the star or planet) at which a dark matter particle will be captured is therefore
\begin{align}
  v_\text{max} = \frac{v_{e}}{\prod_j (1 - z \beta^+_j)^{\tau_j/2}} \,.
  \label{eq:vmax}
\end{align}
Integrating the dark matter velocity distribution $f_*(\vec{v})$ from the minimum dark matter velocity at the stellar body's surface (which is, of course, its escape velocity $v_e$), to $v_\text{max}$ then yields the percentage $P_\text{cap}$ of dark matter particles captured:
\begin{align}
  P_\text{cap} = \int_{v_{e}}^{v_\text{max}} \! \diff v
                 \int_{-1}^{1} \! \diff\!\cos\phi \, f(v) \,.
  \label{eq:Pcap}
\end{align}
Finally, to find the total mass $M_\text{cap}$ of all dark matter particles captured over the lifetime $\mathcal{T}_*\simeq \SI{e9}{yrs}$ of the stellar body, we multiply $P_\text{cap}$ by $\mathcal{T}_*$, by the geometric cross-section $4 \pi R_*^2$, by the local dark matter density $\rho_\chi \simeq \SI{0.3}{GeV/cm^3}$~\cite{Lisanti:2016jxe, Buch:2018qdr, Iocco:2011jz}, and by the average velocity $\langle v_\chi \rangle \simeq \SI{320}{km/s}$ of dark matter particles entering the stellar body. We then integrate over all possible angles $\theta$ to find
\begin{align}
  M_\text{cap} = 4\pi R_*^2 \, \langle v_\chi\rangle \rho_\chi \mathcal{T}_* \,
                 \int_0^{\pi/2} \! 2\sin(\theta)\cos(\theta) P_\text{cap} \, \diff\theta \,.
  \label{eq:Mcap}
\end{align}
Here, the factor $2\sin\theta\cos\theta$ accounts for the solid angle normalization of the entry velocities \cite{Bramante:2019fhi}. It takes into account the fraction of the solid angle covered by an angular interval $\diff\theta$ as well as the fact that particles arriving under a shallow angle ($\theta \sim \pi/2$) are spread out over a larger surface area. The integral in \cref{eq:Mcap} must be computed numerically due to the complicated dependence of $P_\text{cap}$ on $\theta$.

\begin{figure}
    \centering
    \includegraphics[width=\linewidth]{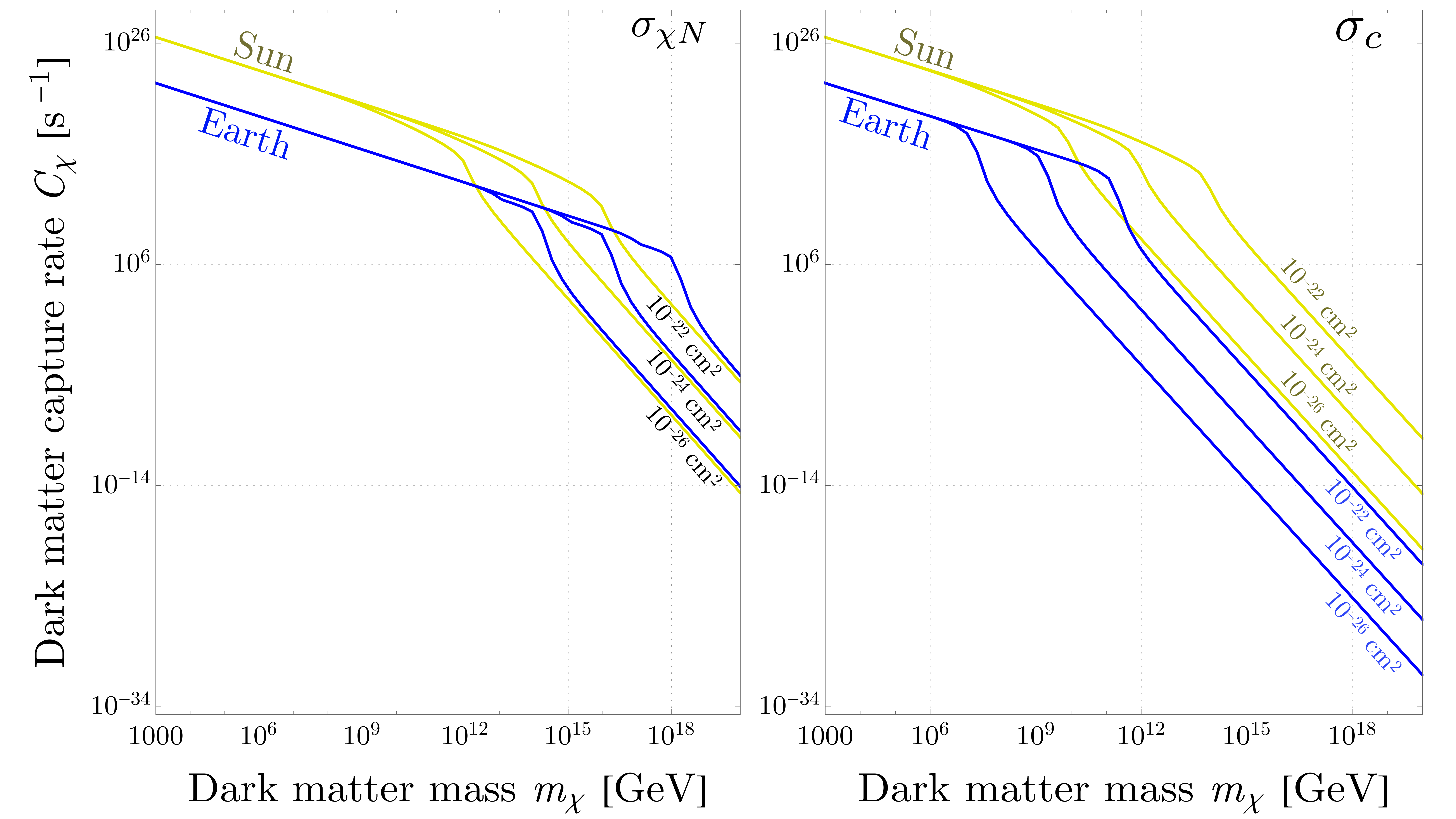}
    \caption{Dark matter capture rates for the two scenarios introduced in \cref{sec:models}, namely isotope-dependent scattering as for standard spin-independent couplings (left panel) and isotope-independent scattering as for some composite asymmetric dark matter models (right panel). Blue lines are for capture in the Earth, while yellow lines correspond to capture in the Sun. 
    In each case, we show results for three different values of $\sigma_{\chi N}$ ($\sigma_{c}$), as indicated in the plot. For the Sun, we have used the stellar $^4$He and $^1$H density profile given in Ref.~\cite{Bahcall:2004pz}, while for the Earth's density and composition we follow Ref.~\cite{Bramante:2019fhi}.}
    \label{fig:CaptureRatePlots}
\end{figure}

The capture rate, defined as the number of dark matter particles captured per unit time, is finally given by
\begin{align}
    C_\chi = \frac{M_\text{cap}}{m_\chi \mathcal{T}_*} \,.
    \label{eq:cchi}
\end{align}
In \cref{fig:CaptureRatePlots}, we plot the dark matter capture rate in the Sun (yellow) and in the Earth (blue) as a function of the dark matter mass $m_\chi$ for various values of the scattering cross-section, $\sigma_{\chi N}$ for isotope-dependent DM--nucleus scattering (left panel) and $\sigma_c$ for isotope-independent dark matter (right panel). We distinguish two qualitatively different regions in these plots: at low $m_\chi$, capture is so efficient that almost all dark matter particles that enter the Sun or Earth are captured. At large $m_\chi$, on the other hand, only a small fraction of the total dark matter flux is swept up. We can approximately fit the curves in \cref{fig:CaptureRatePlots} in these two regimes. The resulting approximate expressions for the capture rates using isotope-dependent ($C_{\chi N}$) and isotope-independent ($C_{c}$) cross-sections in the Earth ($\oplus$) and Sun ($\odot$) are:
\begin{align}
    C_{\chi N}^\oplus &\simeq
    \begin{cases} 
        2.45\times10^{22}\textrm{ s}^{-1}\left( \frac{m_\chi}{10^3\textrm{ GeV}} \right)^{-1}
            & \left(\frac{m_\chi}{1.66\times10^{12}\textrm{ GeV}}\right)<\left(\frac{\sigma_{\chi N}}{10^{-26}\textrm{ cm}^2} \right)
            \\
        8.74\times10^{27}\textrm{ s}^{-1}\left(\frac{m_\chi}{10^8\textrm{ GeV}}\right)^{-7/2}\left(\frac{\sigma_{\chi N}}{10^{-26}\textrm{ cm}^2}\right)^{5/2}
            & \left(\frac{m_\chi}{1.66\times10^{12}\textrm{ GeV}}\right)>\left(\frac{\sigma_{\chi N}}{10^{-26}\textrm{ cm}^2}\right)
    \end{cases} \,,   
    \label{eq:C-fit-1} \\[0.2cm]
    C_c^\oplus &\simeq
    \begin{cases} 
        2.45\times10^{22}\textrm{ s}^{-1}\left( \frac{m_\chi}{10^3\textrm{ GeV}} \right)^{-1}
            & \left(\frac{m_\chi}{2.38\times10^5\textrm{ GeV}}\right)<\left(\frac{\sigma_{c}}{10^{-26}\textrm{ cm}^2}\right)
            \\
            6.8\times10^{10}\textrm{ s}^{-1}\left(\frac{m_\chi}{10^8\textrm{ GeV}}\right)^{-7/2}\left(\frac{\sigma_{c}}{10^{-26}\textrm{ cm}^2}\right)^{5/2}
            & \left(\frac{m_\chi}{2.38\times10^5\textrm{ GeV}}\right)>\left(\frac{\sigma_{c}}{10^{-26}\textrm{ cm}^2}\right)
    \end{cases} \,,          
    \label{eq:C-fit-2} \\[0.2cm]
    C_{\chi N}^\odot &\simeq
    \begin{cases}
        3.45\times10^{26}\textrm{ s}^{-1}\left(\frac{m_\chi}{10^3\textrm{ GeV}}\right)^{-1}
        & \left(\frac{m_\chi}{2.16\times10^{10}\textrm{ GeV}}\right)<\left(\frac{\sigma_{\chi N}}{10^{-26}\textrm{ cm}^2}\right)
        \\
        7.45\times10^{23}\textrm{ s}^{-1}\left(\frac{m_\chi}{10^9\textrm{ GeV}}\right)^{-7/2}\left(\frac{\sigma_{\chi N}}{10^{-26}\textrm{ cm}^2}\right)^{5/2}
        & \left(\frac{m_\chi}{2.16\times10^{10}\textrm{ GeV}}\right)>\left(\frac{\sigma_{\chi N}}{10^{-26}\textrm{ cm}^2}\right)
    \end{cases} \,,
    \label{eq:C-fit-3} \\[0.2cm]
    C_{c}^\odot &\simeq
    \begin{cases}
        3.45\times10^{26}\textrm{ s}^{-1}\left(\frac{m_\chi}{10^3\textrm{ GeV}}\right)^{-1}
        & \left(\frac{m_\chi}{1.93\times10^8\textrm{ GeV}}\right)<\left(\frac{\sigma_c}{10^{-26}\textrm{ cm}^2}\right)
        \\
        5.66\times10^{18}\textrm{ s}^{-1}\left(\frac{m_\chi}{10^9\textrm{ GeV}}\right)^{-7/2}\left(\frac{\sigma_{c}}{10^{-26}\textrm{ cm}^2}\right)^{5/2}
        & \left(\frac{m_\chi}{1.93\times10^8\textrm{ GeV}}\right)>\left(\frac{\sigma_c}{10^{-26}\textrm{ cm}^2}\right)
    \end{cases} \,.          
    \label{eq:C-fit-4}
\end{align}
An analytic derivation of the scaling behavior observed here is given in \cref{app:cap}.

Comparing the capture rates for isotope-independent and isotope-dependent dark matter scattering, we observe from \cref{fig:CaptureRatePlots} that the two scenarios lead to similar capture rates in the Sun, while in the Earth, capture becomes inefficient at much higher masses for isotope-dependent dark matter (left panel). This can be understood from \cref{eq:crossec}, which shows that the cross-section for isotope-dependent dark matter is significantly enhanced for scattering on heavy nuclei, which are abundant in the Earth, but not in the Sun.

\section{Dark Matter Thermalization}
\label{sec:thermalization}

After being gravitationally captured, a dark matter particle follows a closed orbit that intersects with the stellar object. The subsequent thermalization with the surrounding ordinary matter proceeds in a two-stage process \cite{Kouvaris:2010jy}: initially, dark matter orbits are larger than the size of the stellar object (first thermalization). After enough energy is lost through repeated crossings, the orbits become completely contained within the stellar object and the dark matter particles continue to lose energy as their trajectories shrink to a final thermal radius (second thermalization).  This description is only valid if the DM--nucleus cross-section is not too large. Otherwise, the dark matter particles will rapidly thermalize with the outer layers of the stellar object immediately after being captured, and then drift inefficiently towards the core over a very long timescale, a scenario on which we will comment in \cref{sec:drifttime}.

\subsection{First thermalization}
\label{sec:1st-therm}

At the start of the thermalization process, the recently-captured dark matter particles have orbits larger than the size of the stellar object, crossing it multiple times and depositing a fraction of their kinetic energy in each crossing. As in \cref{eq:losspernuc}, the average energy loss a dark matter particle experiences when scattering off a nucleus of type $j$ is \cite{Kouvaris:2010jy, Acevedo:2019gre}
\begin{align}
  \Delta{E} &\xrightarrow{m_\chi \gg m_j(r)} \frac{2 m_j(r)}{m_\chi} E_\text{kin}
            = \frac{2 m_j(r)}{m_\chi} \big[E - m_\chi \phi(r) \big] \,.
  \label{eq:DeltaE}
\end{align}
In the last equality, we have expressed the dark matter particle's kinetic energy, $E_\text{kin}$, in terms of its total energy, $E$, and the gravitational potential,
\begin{align}
  \phi(r) = -\frac{G M_*}{R_*} + \int_{R_*}^{r} \! \frac{G M(r')}{r'^2} \, \diff r' \,.
  \label{eq:phi}
\end{align}
Here, $M(r) = \int_0^{r} \! 4\pi r'^2 \rho_*(r') \, \diff r'$ is the stellar or planetary mass enclosed in a sphere of radius $r$, numerically computed from the profiles displayed in \cref{fig:DensityTemperature}, and $M_* \equiv M(R_*)$ is the total mass.
In \cref{eq:DeltaE}, we assume for simplicity that, at any given radius $r < R_*$, scattering occurs only on a single species of nuclei, hence the dependence on $r$ in $m_j(r)$.  In the case of the Earth, this most abundant element is \iso{Fe}{56} in the core and \iso{O}{16} in the mantle and crust. We have verified that corrections from scattering on subdominant elements are negligible. They become even less important after taking into account that the cross-section required to complete this first thermalization stage within a given time is completely negligible compared to the requirements coming from the second thermalization stage (see below).

We compute the average energy loss per scatter by integrating \cref{eq:DeltaE} over the size of the stellar body:
\begin{align}
    \langle\Delta{E}\rangle_* = \frac{1}{R_*}\int_0^{R_*} \! 2 m_j(r) \bigg( \frac{E}{m_\chi} - \phi(r) \bigg) \, \diff r \,.
    \label{eq:DeltaE-avg}
\end{align}
The right-hand side of this equation is based on the assumption that the dark matter particle's trajectory carries it along a straight line through the very core of the star or planet. This is consistent with the fact that a significant fraction of captured dark matter particles did indeed travel through the core (i.e.\ the region with the highest optical depth) when being captured, so the closed orbit they are scattered into will intersect the core as well.  For other dark matter particles, we expect the approximations in \cref{eq:DeltaE-avg} to introduce an $\mathcal{O}(1)$ error.

To compute the timescale over which first thermalization happens, we need to take into account the orbital period of the captured dark matter particles.  Treating the star or planet as a point-mass for this purpose, the orbital period for a particle with energy $E < 0$ is
\begin{align}
  \Delta{t} = 2 \pi \sqrt{\frac{[r_p(E)]^3}{G M_*}} \,,
  \label{eq:Delta-t}
\end{align}
with the periapsis distance ($i.e.$ the minimum distance to the center of gravity along the orbit) given by
\begin{align}
    r_p(E)  =-\frac{G M_* m_\chi}{E} \,.
    \label{eq:r_p}
\end{align}
The average energy loss rate during the first thermalization phase is obtained by taking the ratio of \cref{eq:DeltaE-avg,eq:Delta-t}, and multiplying by the optical depth $\tau$ to account for the average number of scatters in a single crossing of the star or planet:
\begin{align}
  \frac{\diff E}{\diff t} \simeq -\frac{\tau \langle \Delta E \rangle_*}{\Delta{t}}
  \label{eq:dEdt-1st-therm}
\end{align}
It is convenient to express this energy loss rate in terms of the dimensionless variable
\begin{align}
  \epsilon \equiv \frac{E}{m_\chi} \frac{R_*}{G M_*} \,,
\end{align}
so that the periapsis distance becomes $r_p(\epsilon) = -R_* / \epsilon$, and the orbital period turns into
\begin{align}
  \Delta{t} = 2\pi \frac{R_*^{3/2}}{\sqrt{GM_*}} |\epsilon|^{-3/2} \,.
\end{align}
For the energy loss rate, we arrive at
\begin{align}
  \frac{\diff\epsilon}{\diff t} = -\frac{\tau \langle \Delta \epsilon \rangle_*}{\Delta{t}} \,,
  \label{eq:deps-dt}
\end{align}
with
\begin{align}
  \langle \Delta\epsilon \rangle_* = \frac{1}{R_*}\int_0^{R_*} \!
           \frac{2 m_j(r)}{m_\chi} \bigg( \epsilon - \frac{R_* \phi(r)}{GM_*} \bigg) \, \diff r \,.
  \label{eq:Delta-eps-avg}
\end{align}
\Cref{eq:dEdt-1st-therm} or \cref{eq:deps-dt} must be numerically integrated from an initial energy 
\begin{align}
  E_i = -\frac{2m_j}{m_\chi} E_\text{kin}(R_*) \approx -2 m_j \frac{GM_*}{R_*} \lesssim 0 \,
\end{align}
($\epsilon_i = -2 m_j/m_\chi$), corresponding to the dark matter scattering once at the surface after being captured, to a final energy $E_f = -G M_* m_\chi/R_*$ ($\epsilon_f = -1$) given by the binding energy at the surface. For energies smaller than $E_f$, the dark matter will be completely contained inside the stellar object, and its subsequent evolution is described by the second thermalization stage, discussed in the next section.

We now compute the duration of the first thermalization stage in the case of capture in the Sun. In this case, $m_j(r) \equiv m_j \approx \SI{0.93}{GeV}$ throughout the star.  We find
\begin{align}
  t_\text{th}^{(I)} &= \frac{\pi R_{\odot}^{3/2} m_\chi}{\tau_{\odot} m_j \sqrt{GM_{\odot}}}
                      \int_{\frac{2 m_j}{m_\chi}}^{1} \frac{\diff\epsilon}{\epsilon^{3/2}(2.38 - \epsilon)}
                                                                 \notag\\
                    &\simeq \SI{4.9e7}{yrs} \times \bigg( \frac{m_\chi}{\SI{e7}{GeV}} \bigg)^{3/2}
                                                   \bigg( \frac{\SI{e-54}{cm^2}}{\sigma_{\chi N}} \bigg) \,,
  \label{eq:tth}
\end{align}
where the second equality holds in the limit $m_\chi \gg m_p$.  We have expressed the optical depth in terms of the DM--nucleon scattering cross-section as $\tau_\odot \approx \sigma_{\chi N}/(\SI{e-34}{cm^2})$. The numerical factor in the integrand corresponds to the radial average of the gravitational potential $\phi(r)$, i.e.\ the second term in \cref{eq:Delta-eps-avg}. If we had assumed a homogeneous density profile, this numerical factor would have evaluated exactly to $4/3$, see e.g.\ \cite{Kouvaris:2010jy,Acevedo:2019gre}.

Within the parameter ranges considered here, we find the DM--nucleon scattering cross-section required for the first stage of thermalization to occur within a given time to be much smaller than the one required to form a self-gravitating sphere, as we will see shortly.  In other words, first thermalization is not the limiting factor in the formation of asymmetric dark matter black holes in the Earth and in the Sun.  The minimum DM--nucleon cross-sections required for the dark matter to complete the first thermalization stage within \SI{1}{yr} for the Sun and the Earth are, respectively,
\begin{align}
  \sigma_{\chi N}^{\odot}  &\gtrsim \SI{e-47}{cm^2} \, \bigg( \frac{m_\chi}{\SI{e7}{GeV}} \bigg)^{\frac{3}{2}} \,,
                           & \text{(Sun)}
                              \label{eq:sigma-1st-th-Sun} \\
  \sigma_{\chi N}^{\oplus} &\gtrsim \SI{e-52}{cm^2} \, \bigg( \frac{m_\chi}{\SI{e7}{GeV}} \bigg)^{\frac{3}{2}} \,.
                           & \text{(Earth)}
                              \label{eq:sigma-1st-th-Earth}
\end{align}
We will see below that these cross-sections are much smaller than even the ones required to form a self-gravitating dark matter core within \SI{1}{Gyr} (cf.\ \cref{fig:dmnucleon} below). The estimates from \cref{eq:sigma-1st-th-Sun,eq:sigma-1st-th-Earth} hold for isotope-dependent dark matter--nucleus cross-sections. For isotopes-independent cross-sections, the constraint on $\sigma_{\chi N}^{\odot}$ remains essentially unchanged, while the one on $\sigma_{\chi N}^{\oplus}$ is larger by about seven orders of magnitude. Even in this case, it is still negligible compared to the self-gravitation requirement derived below, though.

\subsection{Second thermalization}
\label{sec:2nd-therm}

Once dark matter particles have lost sufficient energy from successive star or planet crossings for their orbits to become completely contained within the stellar object, they enter the second stage of thermalization. They continue to lose energy through repeated scatterings against nuclei, but the theoretical treatment of this energy loss needs to be somewhat different from the discussion in \cref{sec:1st-therm}. A dark matter particle entering the second thermalization stage still moves on average much faster than the nuclei it scatters against. However, its final thermal velocity will be much smaller compared to the thermal velocity of the nuclei, i.e.\ $\sqrt{3T/m_\chi} \ll \sqrt{3T/m_j}$ since $m_\chi \gg m_j$. In other words, dark matter particles start the second thermalization stage in an ``inertial'' regime where $v_\chi \gg v_j$, but complete it in a ``viscous'' regime where $v_\chi \ll v_j$ \cite{Janish:2019nkk, Acevedo:2019gre}. 

In the inertial regime, energy is lost at a rate that is completely dominated by the dark matter velocity,

\begin{align}
    \frac{\diff E}{\diff t} \bigg|_\text{inertial} &\simeq -\rho_j \sigma_{\chi j} v_{\chi}^3 \,,
                                         &\text{(inertial regime)}
    \label{eq:dEdt-inertial-1}                                  
    \intertext{In the viscous regime, in contrast, energy is lost at a much lower rate given by }
    \frac{\diff E}{\diff t} \bigg|_\text{viscous} &\simeq -\rho_j \sigma_{\chi j} v_\text{th}^{(j)} v_\chi^2 \,.
                                        &\text{(viscous regime)}
    \label{eq:dEdt-viscous-1}
\end{align}
More details on the derivation of these equations are given in \cref{app:collapse}. The hierarchy between these two rates can be appreciated by considering that $v_\chi$ for a particle just entering the second thermalization stage is comparable to the escape velocity at the surface, so that $v_\chi^3 \sim (2 G M_\odot / R_\odot )^{3/2} \sim 10^{-8}$ for the Sun. Close to completion of thermalization in the viscous regime, on the other hand, the average dark matter speed is $v_\chi \sim \sqrt{3T / m_\chi}$, and $v_\text{th}^{(j)} v_\chi^2 \sim 10^{-15} \times (\SI{e7}{GeV} / m_\chi)$.  A similar hierarchy is obtained in the case of the Earth.  In view of this hierarchy, we conservatively compute the second thermalization time using the energy loss rate in the viscous regime.  We make the replacement $v_\chi^2 = (2/m_\chi) (E - \phi(0))$ in \cref{eq:dEdt-viscous-1}, where $E$ is the total energy of the dark matter particle and we have approximated its potential energy by the value of the gravitational potential at its minimum (at $r=0$), which is of order $\sim -0.2 G M_* m_\chi / R_*$.  We thus have to solve
\begin{align}
  \frac{\diff E}{\diff t} \simeq -\Gamma_\text{therm} \times (E-\phi(0)) \,,
  \label{eq:dEdt-viscous-2}
\end{align}
with the characteristic thermalization rate
\begin{align}
  \Gamma_\text{therm} \equiv \frac{2\rho_j \sigma_{\chi j} v_\text{th}^{(j)}}{m_\chi} \,.
  \label{eq:Gamma-therm}
\end{align}
The general solution to \cref{eq:dEdt-viscous-2} is
\begin{align}
  E(t) = \phi(0) + (E_i - \phi(0)) \ e^{-\Gamma_\text{therm} t} \,,
\end{align}
where the initial energy of the dark matter particle is $E_i = \phi(0) + (m_\chi / 2 m_j) T$, corresponding to a dark matter particle moving as fast as the surrounding nuclei. The final energy when thermalization is complete will be $E_f = \phi(0) + 3 T / 2$. This sets the timescale for second thermalization,
\begin{align}
  t_\text{th}^{(II)} = \Gamma_\text{therm}^{-1} \log\left[\frac{E_i - \phi(0)}{E_f - \phi(0)} \right]
                     = \frac{m_\chi}{2 \rho_j \sigma_{\chi j} v_\text{th}^{(j)}}
                       \log\left[ \frac{m_\chi}{3 m_j} \right] \,.
    \label{eq:t-th-2}
\end{align}
As we will see below in \cref{sec:collapse-timescale}, a very similar expression also sets an upper bound on the timescale for the dark matter sphere to collapse after it has amassed enough particles to be self-gravitating (unless quantum degeneracy pressure hinders the collapse, in which case the collapse timescale will be longer, see \cref{sec:collapse-timescale}).

For completeness, we note that the preceding treatment has not considered the possibility of dark matter losing energy by exciting collective oscillation modes in nuclear material, either in the solar plasma or in the Earth's solid inner core. This is because the smallest momentum transfer scale involved in thermalization is much larger than the inverse nuclear separation in both cases. For instance, in the Sun, the smallest momentum exchange is set by $q \approx m_j v_\text{th}^{(j)} \sim \SI{1}{MeV}$, which is much larger than the inverse particle separation $n_j^{1/3} \lesssim \SI{5e8}{cm^{-1}} \sim \SI{10}{keV}$ at the core. Therefore, elastic DM--nucleus scattering is the main energy loss channel.

\subsection{Viscous drag force}
\label{sec:drifttime}

We have discussed in the previous sections the capture and thermalization rates for dark matter particles as a function of their cross-section and mass. For the first and second thermalization descriptions to be valid, however, we required the DM--nucleus cross-section to be sufficiently low for the mean free path of captured dark matter particles to be sizeable. This is necessary for the dark matter to settle down in a thermalized and virialized state near the core of the stellar object. If, on the other hand, the dark matter scattering cross-section on nuclei is very large, all captured dark matter particles will slow down rapidly already in the outer shells of the Earth or Sun and only drift slowly towards the core under the influence of a viscous drag force. Such high cross-section dark matter will not necessarily form black holes faster than dark matter with a lower scattering cross-section because the time it takes them to drift towards the core can be much longer~\cite{Bramante:2019fhi, Starkman:1990nj, Gould:1989gw, Mack:2007xj}. By computing this drift time and requiring that it be sufficiently short, we determine the maximum DM--nucleus cross-section that can result in black hole formation in the Sun or Earth today.

Following refs.~\cite{Bramante:2019fhi, Starkman:1990nj, Gould:1989gw, Mack:2007xj}, we start from the equilibrium condition between the gravitational and viscous drag forces,
\begin{align}
  \frac{G M(r) m_\chi}{r^2} = v_{\chi} \bigg[ \sum_j n_j(r) \, m_j
                    \big\langle \sigma_{\chi j} v^{(j)}_\text{th}(r) \big\rangle \bigg] \,,
  \label{eq:Fg_v_Fd}
\end{align}
where $M(r)$ is once again the mass contained within a shell of radius $r$. Note that the viscous drag force can be directly obtained by taking the derivative of \cref{eq:dEdt-viscous-1} with respect to $v_\chi$.  Using $v_{\chi} = \partial r / \partial t$ as well as $v_\text{th} \simeq \sqrt{3 T / m_j}$, we integrate \cref{eq:Fg_v_Fd} to estimate the drift time as a function of the cross-section:
\begin{align}
  t_\text{drift} = \frac{1}{G m_\chi} \Bigg[
                     \sum_j \sigma_{\chi j} \int_0^{R_*} \frac{n_j(r) \sqrt{3 m_j T(r)}}{M(r)} r^2 \diff r \bigg] \,.
\label{eq:sigmaDrift}
\end{align}
Setting $t_\text{drift} \simeq \SI{e9}{yrs}$, we can solve for the maximum cross-sections at which the captured dark matter could collapse into black holes:
\begin{align}
  \sigma_{\chi N}^\odot  &\lesssim \SI{e-16}{cm^2} \left( \frac{m_\chi}{\SI{e7}{GeV}} \right)
                         &\text{for the Sun, and}
  \label{eq:tdriftS} \\[0.2cm]
  \sigma_{\chi N}^\oplus &\lesssim \SI{e-21}{cm^2} \left( \frac{m_\chi}{\SI{e7}{GeV}} \right)
                         &\text{for the Earth.}
  \label{eq:tdriftE}
\end{align}
Analogous estimates for isotope-independent dark matter scattering are trivially obtained by replacing $\sigma_{\chi j} \to \sigma_c$.

\section{Gravitational Collapse and Black Hole Formation}
\label{sec:collapse}

After the dark matter particles have completely thermalized at the core of the stellar object, they settle in a small thermal sphere, with a radius which can be estimated according to the virial theorem, $2 \ev{E_\text{kin}} = -\ev{V}$, where we set $\ev{E_\text{kin}} = \tfrac{3}{2} T_*$ and $\ev{V} = -\tfrac{4}{3} \pi r_\text{th}^2 \rho_* G m_\chi$. For the purpose of this estimate, we assume the dark matter density $\rho_\chi$ to be smaller than the density $\rho_*$ of the host, so that the contribution of the dark matter to the total gravitational potential is negligible. This leads to 
\begin{align}
  r_\text{th} = \sqrt{\frac{9 T_*}{4 \pi G \rho_* m_\chi}}
              &\approx \SI{30}{km} \times \bigg( \frac{\SI{e7}{GeV}}{m_{\chi}} \bigg)^{\frac{1}{2}} 
                                          \bigg( \frac{T_\odot}{\SI{1.5e7}{K}} \bigg)^{\frac{1}{2}}
                                          \bigg( \frac{\SI{156}{g/cm^3}}{\rho_\odot} \bigg)^{\frac{1}{2}} \notag\\
              &\approx \ \SI{2}{km} \times \bigg( \frac{\SI{e7}{GeV}}{m_{\chi}} \bigg)^{\frac{1}{2}} 
                                           \bigg( \frac{T_\oplus}{\SI{5e3}{K}} \bigg)^{\frac{1}{2}}
                                           \bigg( \frac{\SI{10}{g/cm^3}}{\rho_\oplus} \bigg)^{\frac{1}{2}} \,.
  \label{eq:rth}
\end{align}
The two expressions on the right-hand side have been normalized to the densities and temperatures corresponding to the solar and terrestrial core, respectively. In both cases, the thermal radius is much smaller than the respective inner core size.

The small dark matter sphere grows rapidly in mass as more and more non-annihilating dark matter is captured and thermalizes, eventually becoming gravitationally unstable. Once this gravitational instability is reached, it collapses and forms a black hole at the center which, depending on its initial mass, can either evaporate or grow through accretion of stellar or planetary material and captured dark matter. In the following we discuss the specific conditions that must be met for the dark matter sphere to collapse, and we calculate the time evolution of the resulting black hole.

\subsection{Conditions for Dark Matter Collapse}
\label{sec:collapse-conditions}

The dark matter sphere will become unstable and collapse provided the following two conditions are met:
\begin{enumerate}
  \item \textit{Jeans instability}:
    the sound crossing time ($t_s$) must be greater than the dark matter free-fall time ($t_\text{ff}$).    The sound crossing time depends on the sound speed, which we take to be $c_s \approx \sqrt{T / m_\chi}$ (neglecting dark matter self-interactions), so that
    \begin{align}
      t_s = \frac{r_\text{th}}{c_s}
          = \frac{3}{\sqrt{4\pi G \rho_*}}
          &\approx  \SI{260}{s} \times \bigg( \frac{\SI{156}{g/cm^3}}{\rho_{\odot}}  \bigg)^\frac{1}{2} \notag\\
          &\approx \SI{1030}{s} \times \bigg( \frac{ \SI{10}{g/cm^3}}{\rho_{\oplus}} \bigg)^{\frac{1}{2}} \,.
      \label{eq:ts}
    \end{align}
    Here, we have used $r_\text{th}$ from \cref{eq:rth}, and the two scaling laws on the right row correspond once again to dark matter captured in the Sun and in the Earth, respectively. The dark matter free-fall time is most easily computed by considering the free-fall of a test mass onto an object of mass $\tfrac{4}{3} \pi \rho_\chi r_\text{th}^3$.  This free-fall can be viewed as the limiting case of an orbital motion with eccentricity $e \to 1$ and semi-major axis $r_\text{th}/2$.  By invoking Kepler's third law to compare to a circular orbit with the same semi-major axis, the free-fall time is readily obtained as
    \begin{align}
      t_\text{ff} = \sqrt{\frac{3 \pi}{32 G \rho_\chi}}
                    \approx \SI{167}{s} \times \bigg( \frac{\SI{156}{g/cm^3}}{\rho_{\chi}} \bigg)^{\frac{1}{2}}
                    \approx \SI{659}{s} \times \bigg( \frac{ \SI{10}{g/cm^3}}{\rho_{\chi}} \bigg)^{\frac{1}{2}} \,.
      \label{eq:tff}
    \end{align}
    Comparing \cref{eq:ts,eq:tff}, we see that the Jeans instability condition $t_\text{ff} \lesssim t_s$ is therefore satisfied when the dark matter sphere reaches a density larger than the solar or terrestrial core density. 
    
    It is also instructive to consider the DM--nucleus and DM--DM interaction timescales: they will determine the processes by which dark matter sheds energy and angular momentum while collapsing. The DM--nucleus interaction time is
    \begin{align}
        t_{\chi j} = (n_j \sigma_{\chi j} v_\text{th}^{(j)})^{-1}
                   &\approx\ \SI{300}{s} \times \bigg( \frac{\SI{e-36}{cm^2}}{\sigma_{\chi j}} \bigg)
                                                \bigg( \frac{\SI{1.5e7}{K}}{T} \bigg)^{\frac{1}{2}} \notag\\
                   &\approx\ \SI{1.2e5}{s} \times \bigg( \frac{\SI{e-36}{cm^2}}{\sigma_{\chi j}} \bigg)
                                                \bigg( \frac{\SI{5e3}{K}}{T} \bigg)^{\frac{1}{2}} \,,
    \intertext{whereas the DM--DM interaction time is}
        t_{\chi\chi} = (n_\chi \sigma_{\chi\chi} v_\chi)^{-1}
                     &\approx\ \SI{100}{s} \times \bigg( \frac{\SI{e-25}{cm^2}}{\sigma_{\chi\chi}} \bigg)
                                                  \bigg( \frac{m_\chi}{\SI{e7}{GeV}} \bigg)^{\frac{3}{2}}
                                                  \bigg( \frac{\SI{156}{g/cm^3}}{\rho_\chi} \bigg)
                                                  \bigg( \frac{\SI{1.5e7}{K}}{T} \bigg)^{\frac{1}{2}} \notag\\
                     &\approx\ \SI{8.5e4} {s} \times \bigg( \frac{\SI{e-25}{cm^2}}{\sigma_{\chi\chi}} \bigg)
                                                  \bigg( \frac{m_\chi}{\SI{e7}{GeV}} \bigg)^{\frac{3}{2}}
                                                  \bigg( \frac{\SI{10}{g/cm^3}}{\rho_\chi} \bigg)
                                                  \bigg( \frac{\SI{5e3}{K}}{T} \bigg)^{\frac{1}{2}} \,.
    \end{align}
    Note that $t_{\chi\chi}$ has a dependence on $m_\chi$ (stemming from the rewriting of $n_\chi$ in terms of $\rho_\chi$ and from $v_\chi \sim \sqrt{3 T/m_\chi}$), while $t_{\chi j}$ does not.
    
    In future work, it may be interesting to consider scenarios in which dark matter emits light dark sector particles while collapsing, and to investigate whether these dark sector particles themselves can be detected. We discuss a few aspects of this here, although this will not affect our main analysis, in which DM--nuclear interactions account for dark matter energy loss during collapse. In the case that dark matter energy loss also occurred from the emission of light dark sector particles, if $t_{\chi\chi}$ is small enough compared to $t_\text{ff}$, the collapsing dark matter will form an isothermal sphere that would radiate like a blackbody. Comparing $t_{\chi\chi}$ to $t_\text{ff}$, we see that a self-interaction cross-section $\gg \SI{e-25}{cm^2}$ is required for dark matter to self-thermalize as it collapses.

  \item \textit{Critical mass}:
    We have determined that prior to collapse the dark matter particles are in thermal equilibrium at temperature $T_*$ (the internal temperature of the star or planet) and form a virialized sphere of radius $r_\text{th}$, see \cref{eq:rth}. The condition leading to \cref{eq:rth}, however, needs to be refined when the mass density $\rho_\chi$ of dark matter within radius $r_\text{th}$ becomes comparable to the density $\rho_*$ of ordinary matter. In this case, when invoking the virial theorem, we can no longer neglect the gravitational potential generated by the dark matter particles. Instead, we have to solve
    \begin{align}
        \frac{3}{2} T_* = \frac{1}{2} G \bigg( \frac{4}{3} \pi \rho_* m_\chi r_\text{th}^2
                                                + \frac{M_\text{cap} m_\chi}{r_\text{th}} \bigg) \,.
        \label{eq:Msg-condition}
    \end{align}
    This equation has a real solution for $r_\text{th}$ only if the total mass of captured dark matter, $M_\text{cap}$, is below a critical value, which we call the self-gravitating mass $M_\text{sg}$. 
    \begin{align}
        M_\text{sg} = \sqrt{\frac{3 T_*^3}{\pi G^3 m_\chi^3 \rho_*}}
                    &\approx \SI{3e45}{GeV} \times \bigg( \frac{\SI{e7}{GeV}}{m_\chi} \bigg)^{\frac{3}{2}}
                                                   \bigg( \frac{T_\odot}{\SI{1.5e7}{K}} \bigg)^{\frac{3}{2}}
                                                   \bigg( \frac{\SI{156}{g/cm^3}}{\rho_\odot} \bigg)^{\frac{1}{2}}  \notag\\
                    &\approx \SI{8e40}{GeV} \times \bigg( \frac{\SI{e7}{GeV}}{m_\chi} \bigg)^{\frac{3}{2}}
                                                   \bigg( \frac{T_\oplus}{\SI{5e3}{K}} \bigg)^{\frac{3}{2}}
                                                   \bigg( \frac{\SI{10}{g/cm^3}}{\rho_\oplus} \bigg)^{\frac{1}{2}} \,,
        \label{eq:sg}
    \end{align}
    If $M_\text{cap} > M_\text{sg}$, \cref{eq:Msg-condition} indicates that the dark matter core can no longer be stabilized by thermal pressure and will collapse.
    
    In addition to satisfying $M_\text{cap} > M_\text{sg}$, the dark matter sphere must also be sufficiently massive so that it cannot be stabilized at a smaller radius than $r_\text{th}$ by dark matter self-interactions or degeneracy pressure. Here, we consider the maximum mass that can be stabilized by Fermi degeneracy pressure for non-interacting fermions,
    \begin{align}
        M_\text{f} \sim \frac{\mpl^3}{m_\chi^2}
                   \approx \SI{2e43}{GeV} \times \bigg(   \frac{\SI{e7}{GeV}}{m_\chi} \bigg)^2 \,,
        \label{eq:chandraMass}
    \end{align}
    where $\mpl = G^{-1/2}$ is the Planck mass. This estimate aligns well with more detailed modeling of quantum degeneracy pressure in spheres composed of fermionic dark matter \cite{Gresham:2018rqo}. If the dark matter consists of composite objects of mass $m_\chi$, made up of constituent fermions of mass $m_f$, then $m_\chi$ in \cref{eq:chandraMass} needs to be replaced by $m_f$. The estimate from \cref{eq:chandraMass} applies also in the case of bosonic dark matter with an order-unity quartic self-coupling~\cite{Colpi:1986ye}.
    
    We therefore define the critical mass required for collapse as the maximum between the self-gravitating mass and the maximum stable mass, 
    \begin{align}
        M_\text{crit} = \max \left[ M_\text{sg}, M_\text{f} \right] \,.
        \label{eq:Mcrit}
    \end{align}
\end{enumerate}
Comparing the Jeans instability condition and the critical mass condition, we find that the latter condition is the more stringent one. The reason is that the collapse condition $M_\text{cap} > M_\text{sg}$ is roughly equivalent to the requirement that mass density of dark matter, $\rho_\chi$, be larger than the mass density of ordinary matter, $\rho_*$ (up to $\mathcal{O}(1)$ factors). In parameter regions where $M_\text{crit} = M_\text{sg}$ (i.e.\ where degeneracy pressure is irrelevant), the Jeans instability and critical mass conditions are therefore equivalent. If $M_\text{crit} = M_\text{f}$ (collapse inhibited by degeneracy pressure), the critical mass condition is more stringent. Therefore, we will in the following only consider the condition that a total dark matter mass larger than $M_\text{crit}$ is collected as the necessary condition for collapse and black hole formation.

\subsection{Collapse timescale}
\label{sec:collapse-timescale}

Once the critical amount of dark matter mass has been collected, collapse of the now unstable dark matter sphere proceeds and dark matter particles shed their gravitational energy by repeated scattering on nuclei.  As discussed above, the dark matter particles are, at the time of collapse, moving much slower than the nuclei, so their energy loss during collapse is initially described by the viscous regime, in particular by \cref{eq:dEdt-viscous-1} (see also \cref{app:collapse}). The total transferred energy corresponds to the gravitational energy released during collapse. The initial kinetic energy of a dark matter particle in the thermal sphere is of order $\ev{E_i} = G M_\text{crit} m_\chi / (2 r_\text{th})$ by virtue of the virial theorem. The dark matter particle exits the viscous regime when $v_\chi \sim \sqrt{3 T / m_j}$, or in terms of kinetic energy, $\ev{E_f} = 3 m_\chi T / (2 m_j)$. Due to the virial theorem its gain in kinetic energy, $\ev{E_f} - \ev{E_i}$, corresponds to an equal loss in total energy. Therefore, we can obtain the time the particle spends in the viscous regime, $t_\text{col}$, by integrating \cref{eq:dEdt-viscous-1} between $\ev{E_i}$ and $\ev{E_f}$. We obtain:
\begin{align}
  t_\text{col} = \frac{m_\chi}{2\rho_j \sigma_{\chi j} v_\text{th}^{(j)}}    
                 \log\bigg[\big[ v_{th}^{(j)} \big]^2
                 \frac{r_\text{th}}{2 G M_\text{crit}} \bigg] \,.
  \label{eq:tcol}
\end{align}
This timescale is close to the total collapse time, since in the subsequent `inertial' regime ($v_\chi \gg \sqrt{3 T/m_j}$), dark matter particles lose energy at a much faster rate. In other words, the collapse timescale is determined by the early stage of collapse in which the DM--nucleus relative velocity is dominated by the thermal motion of the nuclei.

Note also that \cref{eq:tcol} is nearly identical to the timescale of second thermalization, \cref{eq:t-th-2} when $M_\text{crit} = M_\text{sg}$, that is when quantum degeneracy pressure is unimportant. This is not surprising given that both equations are based on integrating the same expression for $\diff E/\diff t$, and only differ slightly in the choice of one of the integration boundaries, which affects the argument of the log by an $\mathcal{O}(1)$ factor.

Before proceeding, we comment on potential quantum mechanical corrections to the purely classical estimates presented so far. Quantum mechanics is relevant when the de~Broglie wavelength $\lambda_\text{db} = (m_\chi v_\chi)^{-1}$ of the dark matter particles becomes comparable to their mean separation.  However, this is never the case: the dark matter particles contained in a radius $r$ in the Solar core have a de~Broglie wavelength $\lambda_\text{db} \sim \SI{e-14}{cm} \times (r/r_\text{th})^{1/2} \times (m_\chi/\SI{e7}{GeV})^{-1/2}$, whereas their mean separation is of order $n_\chi^{-1/3} \sim \SI{e-7}{cm} \times (r/r_\text{th}) \times (m_\chi / \SI{e7}{GeV})^{5/2}$. Comparing both scales, we verify that $\lambda_\text{db} \ll n_\chi^{-1/3}$ for any radius $r \gtrsim 2GM_\text{crit}$ and any value of $m_\chi$ we consider. We arrive at a similar conclusion in the case of the Earth, although the scaling with $r$ is different since the critical mass can be larger than the self-gravitating mass in some portions of the parameter space. We conclude that quantum mechanical effects need not be accounted for in this analysis.

\subsection{Black hole evolution}
\label{sec:BHevol}

Once a black hole has formed, its time evolution is determined by the competing effects of accretion~\cite{Bondi:1944jm} and Hawking radiation~\cite{Hawking:1974rv}, and the rate at which its mass, $\mbh$, changes is given by \cite{Acevedo:2019gre} 
\begin{align}
  \frac{\diff \mbh}{\diff t} = \frac{4 \pi \rho_*{(G \mbh)^{2}}}{c_{s*}^{3}}
                          + e_\chi m_\chi C_\chi
                          - \frac{f(\mbh)}{ (G \mbh)^{2}} \,.
  \label{eq:bhtotal}
\end{align}

The first term on the right hand side of this equation describes Bondi accretion of ordinary matter, with $c_{s*}$ being the sound speed at the core of the stellar object, $c_\odot \approx \sqrt{T/m_p} \approx \SI{350}{km/s}$ for the Sun and $c_\oplus \approx \SI{10}{km/s}$~\cite{Dziewonski:1981xy} for the Earth.
Simple estimates indicate that Bondi accretion should accurately describe the accumulation of baryonic matter in the Earth and Sun onto the black hole. In general, spherical accretion can depend on the angular momentum of the infalling material, which in principle could establish an orbital trajectory, if its initial angular momentum is sizable. However, we find that the angular momentum of the baryonic matter in the Earth and Sun is small enough that it should not affect spherical Bondi accretion onto black holes formed from collapsed dark matter. Earth's specific angular momentum, $h_*$, at the black hole's Bondi radius, $r_{B}=2GM_\text{crit}/c_{s*}^2$ \cite{Bondi:1944jm,Proga:2002ey}, can be estimated using the rotational frequency of the host's material at the Bondi radius, $\mathcal{F}_*$. For $\mathcal{F}_\odot=1667$ nHz and $\mathcal{F}_\oplus=12$ $\mu$Hz for the Sun \cite{refId0} and Earth \cite{1996Natur.382..221S} respectively,
\begin{align}
        h_* \approx 2\pi r_B^2 \mathcal{F}_*
                    &\approx \SI{1.3e-16}{cm} \times \bigg( \frac{\mathcal{F}_\odot}{1667\text{ nHz}} \bigg) \bigg( \frac{\SI{e7}{GeV}}{m_\chi} \bigg)^{3}
                                                   \bigg( \frac{T_\odot}{\SI{1.5e7}{K}} \bigg)^{3}
                                                   \bigg( \frac{\SI{156}{g/cm^3}}{\rho_\odot} \bigg)  \notag\\
                    &\approx \SI{1e-18}{cm} \times \bigg( \frac{\mathcal{F}_\oplus}{12\text{ }\mu\text{Hz}} \bigg) \bigg( \frac{\SI{e7}{GeV}}{m_\chi} \bigg)^{3}
                                                   \bigg( \frac{T_\oplus}{\SI{5e3}{K}} \bigg)^{3}
                                                   \bigg( \frac{\SI{10}{g/cm^3}}{\rho_\oplus} \bigg) \,
    \label{eq:specAM}
\end{align}

\noindent when $M_\textrm{crit}$ is set by the self-gravitation condition. This should be compared to a relativistic specific angular momentum at the black hole's Schwarzschild radius, $h_\textrm{bh} = r_\textrm{sch} c=2GM_\text{crit} c$ \cite{1916SPAW.......189S}, which evaluates to

    \begin{align}
        h_\text{bh}
                    &\approx \SI{8e-8}{cm} \times \bigg( \frac{\SI{e7}{GeV}}{m_\chi} \bigg)^{\frac{3}{2}}
                                                   \bigg( \frac{T_\odot}{\SI{1.5e7}{K}} \bigg)^{\frac{3}{2}}
                                                   \bigg( \frac{\SI{156}{g/cm^3}}{\rho_\odot} \bigg)^{\frac{1}{2}}  \notag\\
                    &\approx \SI{2e-11}{cm} \times \bigg( \frac{\SI{e7}{GeV}}{m_\chi} \bigg)^{\frac{3}{2}}
                                                   \bigg( \frac{T_\oplus}{\SI{5e3}{K}} \bigg)^{\frac{3}{2}}
                                                   \bigg( \frac{\SI{10}{g/cm^3}}{\rho_\oplus} \bigg)^{\frac{1}{2}} \,.
        \label{eq:rsch}
    \end{align}
Comparing Eq.~\eqref{eq:specAM} to \eqref{eq:rsch}, we find that in both the Earth and the Sun $h_*\ll h_\textrm{bh}$ for all $m_\chi$ considered, which supports the use of spherical Bondi accretion in this work. Of course, a more accurate treatment of baryonic accumulation on a small black hole can be obtained using numerical computations. Such a treatment has recently been completed for baryonic accumulation onto a small black hole inside a neutron star \cite{East:2019dxt}, which found that despite the neutron star being composed of a very stiff fluid in which hydrodynamic instabilities are prone to forming, even a near maximally-spinning neutron star accreted baryons with a rate matching Bondi accretion. Hence, in slower-spinning, less hydrodynamically active systems with lower baryonic sound speeds, such as the Earth and Sun, the Bondi accretion regime is probably accurate. 

The second term in Eq.~\eqref{eq:bhtotal} corresponds to black hole growth from the infall of additional dark matter. The dynamics of this process will depend on whether the dark matter is self-thermalized (see above), in which case the Bondi accretion rate applies (with the sound speed of the dark matter sphere inserted), or whether individual particle trajectories must be considered \cite{Doran:2005vm}. The details of dark matter accretion are encoded in the efficiency factor $e_\chi$, which we will for simplicity set to unity. Changing $e_\chi$ to smaller values would shift the boundary between the parameter regions in which black holes consume their hosts and the ones in which they evaporate. 

The third term on the right-hand side of \cref{eq:bhtotal} is finally the Hawking evaporation rate. In this term, $f(\mbh)$ is often called the Page factor \cite{Page:1976df, MacGibbon:1991tj, MacGibbon:1990zk}. It depends on the number of Standard Model degrees of freedom radiating from the black hole, which increases with the black hole temperature, and on so-called grey-body corrections that describe the modifications that the initial black-body spectrum suffers as particles propagate out of the black hole's gravitational field. More precisely, the Page factor is given by
\begin{align}
    f(\mbh) &= \sum_i \int_0^\infty \! \diff E \frac{E}{2\pi} \frac{g_i \Gamma_{s_i}(M_\textrm{BH},E) }{e^{E/\tbh}\pm 1} \,,
    \label{eq:page-factor}    
\end{align}
where the plus (minus) sign in the denominator is for fermions (bosons), the sum runs over all particles with mass $m_i < T_\text{BH}$, and $g_i$ is the number of internal degrees of freedom corresponding to particle species $i$. We consider only SM degrees of freedom (as well as the small contribution fro graviton emission) in the sum as additional heavy BSM states will not drastically modify the evolution in the regimes considered here. The factor $\Gamma_{s_i}$ is the aforementioned grey-body factor, which can be expressed as follows:
\begin{align}
    \Gamma_{s_\alpha}(M_\textrm{BH},E) \equiv \frac{E^2 \sigma_{s_\alpha}(M_\textrm{BH},E)}{\pi} \,,
    \label{eq:grey-body}
\end{align}
where $\sigma_{s_\alpha}$ is the absorption cross-section encoding a particle's probability for escaping the gravitational potential of the black hole. This cross-section depends on the particle's spin $s$. It has been calculated in Ref.~\cite{MacGibbon:1990zk}, and extended to massive fermions in Ref.~\cite{Doran:2005vm}.\footnote{See also Ref.~\cite{Lunardini:2019zob} which explores applications of different neutrino mass mechanisms and the effects on the diffuse backgrounds from a population of evaporating primordial black holes.}  We evaluate \cref{eq:page-factor,eq:grey-body} utilizing the publicly available code \texttt{BlackHawk}~\cite{Arbey:2019mbc}. In \texttt{BlackHawk}, the grey-body factors are described by fitting functions that reproduce the known results in the limits of Schwarzschild and Kerr black holes~\cite{MacGibbon:1990zk, Page:1976df}. When running BlackHawk, we include graviton emission, and we assume a spin-zero black hole.

While our final numerical results will include the full Page factor from \texttt{BlackHawk}, it is instructive to set $f(\mbh) = 1 / (15\,360 \pi)$ for a moment, corresponding to a black hole radiating only photons, wit no grey-body corrections. This will allow us to analytically understand the interplay of accretion and evaporation in \cref{eq:bhtotal}. The balance between the three contributions to $\diff \mbh / \diff t$ determines the range of dark matter masses and scattering cross-sections for which the black hole evaporates rather than consuming its host. To compute the evaporation time, it is convenient to shorten our notation by introducing the abbreviations
\begin{align}
  \ell \equiv \frac{4 \pi \rho_* G^2}{c_*^3} \,, \qquad
  a_1 \equiv e_\chi m_\chi C_\chi \,, \qquad
  a_2 \equiv \sqrt{\frac{\rho_*}{960c_*^3} + a_1^2} \,, \qquad
  b^\pm \equiv \sqrt{\frac{a_2 \pm a_1}{2\ell}} \,.
  \label{eq:coeffDef1}
\end{align}
With these definitions, we can write the evaporation time in the relatively simple form
\begin{align}
  \timebh = \frac{b^+}{a_2} \arctan\bigg(\frac{\mbh^\text{init}}{b^+} \bigg) -\frac{b^-}{a_2} \arctanh\bigg(\frac{\mbh^\text{init}}{b^-} \bigg)
 \,.
  \label{eq:tevapFull}
\end{align}

Relative to the Hawking evaporation rate, the Bondi accretion rate is negligible for initial black hole masses $\mbh^\text{init} \equiv \mbh(t=0)  \lesssim \SI{e14}{g}$, which is determined by $M_\text{crit}$ in \cref{eq:Mcrit}.  However, the maximum solar dark matter accretion rate is greater than the black hole radiation rate for black hole masses greater than $\mbh^\text{init} \approx \SI{e10}{g}$. With this in mind, the evaporation time for a black hole including dark matter accretion (but neglecting Bondi accretion), is
\begin{align}
    \timebh =\frac{\mbh^\text{init}}{m_{\chi} C_{\chi}} - \frac{a^2}{(m_\chi C_{\chi})^{3/2}} \arctanh\left( \frac{\sqrt{m_\chi C_\chi} \mbh^\text{init}}{a^2} \right) \,,
 \label{eq:gensol}
\end{align}
where $a^2 = (G \sqrt{15360 \pi})^{-1}$. It can be shown that $\sqrt{m_{\chi} C_{\chi}} \mbh^\text{init} \ll a^{2}$ implies rapid black hole evaporation. In this limit, \cref{eq:gensol} reduces to usual solution obtained when Hawking radiation term dominates alone,
\begin{align}
    \timebh = 5120 \pi G^2 (\mbh^\text{init})^3
            \approx \SI{e12}{yrs} \times
                    \bigg( \frac{\SI{e12}{GeV}}{m_\chi} \bigg)^{\frac{9}{2}}
                    \bigg( \frac{\SI{156}{g/cm^3}}{\rho_\odot} \bigg)^{\frac{3}{2}}
                    \bigg( \frac{T_\odot}{\SI{1.5e7}{K}} \bigg)^{\frac{9}{2}}.
    \label{eq:bh3}
\end{align}
Although for $m_\chi \sim \SI{e12}{GeV}$ the evaporation time will be much larger than the solar lifetime, the strong scaling with $m_\chi$ implies much shorter evaporation times for heavier dark matter.

\begin{figure}
    \centering
    \includegraphics[width=\linewidth]{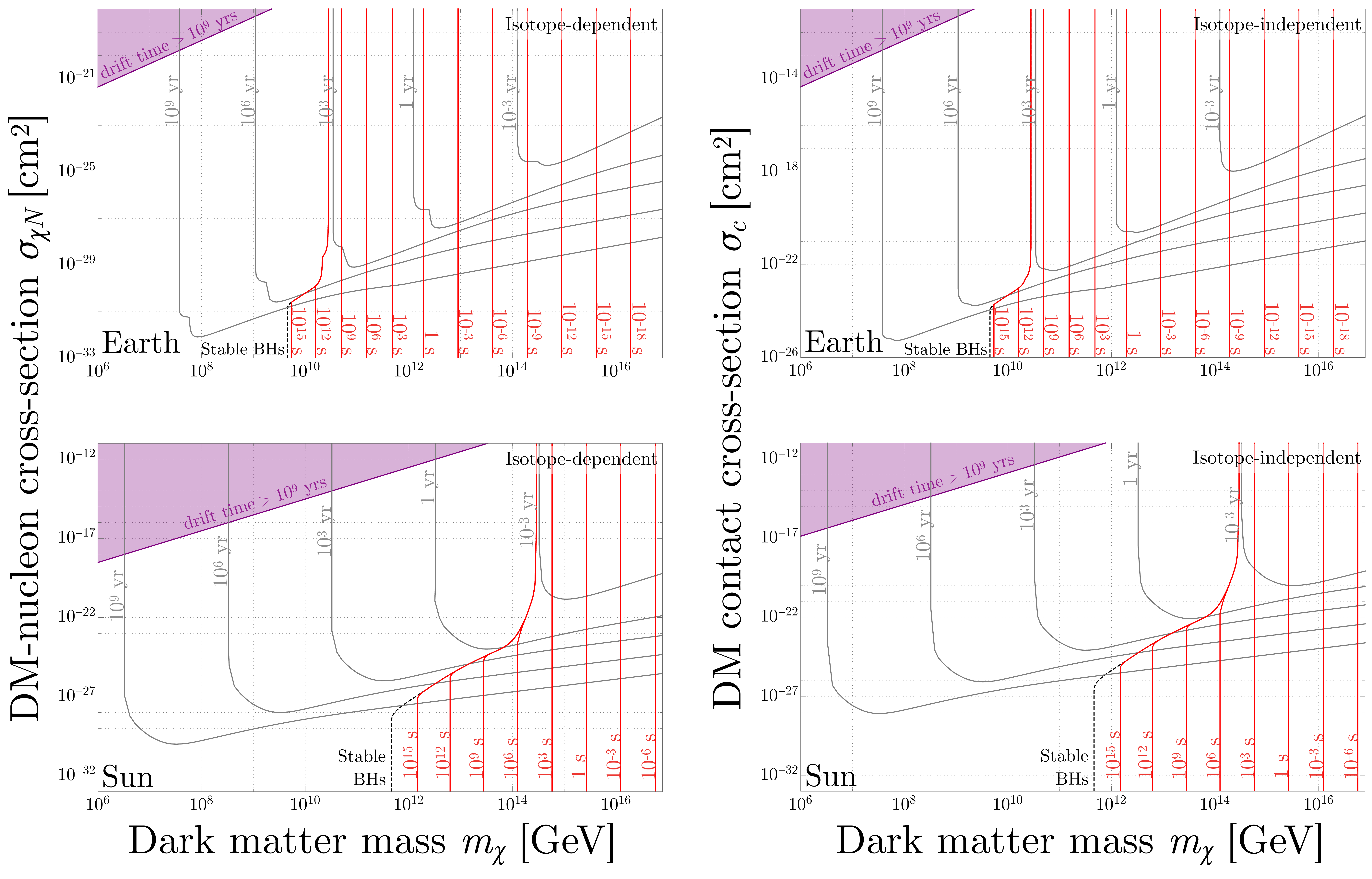}
    \caption{Formation (grey) and evaporation (red) times for the Earth (top panels) and Sun (bottom panels), calculated either with isotope-dependent DM--nucleus cross-sections as for conventional spin-independent couplings (left panels), or with isotope-independent contact cross-sections relevant for instance to some asymmetric composite dark matter models (right panels). The condition that the drift time be $\lesssim1$ Gyr (\cref{eq:tdriftS,eq:tdriftE}) is violated inside the shaded regions at the top-left corner of all plots. The black dashed lines describe stable black holes, in which accretion and evaporation are equal, resulting in a black hole that neither grows nor shrinks. Note that, to make the relevant features more visible, the scale on the vertical axis is not the same in these four panels.}
    \label{fig:tFormEvaps}
\end{figure}

In \cref{fig:tFormEvaps} we compare the time scales for black hole formation (including capture, thermalization, and collapse) and evaporation in the Sun and in the Earth. We moreover compare these time scales in the two scenarios introduced in \cref{sec:models}: spin-independent DM--nucleus scattering with a scattering cross-section that kinematically depends on the mass of the nuclear scattering targets, and isotope-independent DM--nucleus scattering.

We observe that the red evaporation contours in \cref{fig:tFormEvaps} converge at $m_\chi \simeq \SI{e16}{GeV}$ for the Sun ($\simeq \SI{e11}{GeV}$ for the Earth) for evaporation times $\lesssim \SI{1}{hour}$ ($\lesssim \SI{1}{kyr}$). At dark matter masses larger than this boundary, the black holes that form are light enough to evaporate, even when $\mathcal{O}(100\%)$ of the dark matter flux is accreted. In other words, the Hawking radiation rate dominates over both accretion rates in \cref{eq:bhtotal}, and the black hole lifetime is simply given by \cref{eq:bh3}. For lighter dark matter, on the other hand, accretion is faster and the black hole lifetime becomes formally infinite, as can be derived from the full expression for $t_\text{BH}$, \cref{eq:tevapFull}. Note that the exact value of $m_\chi$ at which the red contours converge depends on the dark matter scattering cross-section. This is because for the large cross-sections considered here, accretion of additional dark matter is more important than accretion of ordinary matter, so that the accretion rate is essentially given by the dark matter capture rate. The latter, of course, depends on $\sigma_{\chi N}$ or $\sigma_c$. Note that this dependence disappears above a certain critical value of $\sigma_{\chi N}$ or $\sigma_c$ above which all dark matter particles entering the Sun or Earth are captured. This explains why the boundary between the Hawking radiation regime and the accretion regime becomes independent of the cross-section in the upper regions of the plots in \cref{fig:tFormEvaps}.  Note that the red contours are obviously independent of the capture cross-section for parameter regions in which the black hole lifetime is set by the Hawking evaporation rate.

Regarding the grey contours of constant black hole formation rate in \cref{fig:tFormEvaps}, we observe a characteristic wedge shape. This shape is essentially determined by the dark matter capture rate, which is the bottleneck for black hole formation in most of the parameter space.  The slope of the bottom edge of the wedges can be understood from the scaling behavior of the self-gravitating mass $M_\text{sg}$ and the dark matter capture rate $C_\chi$ with the dark matter mass $m_\chi$ and its scattering cross-section $\sigma = \sigma_{\chi N}$ or $\sigma_c$. Indeed, given that $M_\text{sg} \propto m_\chi^{3/2}$ (see \cref{eq:sg}), $C_\chi \propto \sigma^{5/2} m_\chi^{-7/2}$ (see \cref{eq:C-fit-1,eq:C-fit-2,eq:C-fit-3,eq:C-fit-4}), and the total mass of captured dark matter particles is given by $M_\text{cap} = C_\chi m_\chi$, we readily find that the gray contours in \cref{fig:tFormEvaps} should scale according to $\sigma \propto m_\chi^{2/5}$ in the large-$m_\chi$ limit, where only a small fraction of dark matter particles crossing the Sun or Earth is captured. This scaling is readily confirmed by \cref{fig:tFormEvaps}.  The slope changes towards lower dark matter masses, where also dark matter particles with velocities significantly above the escape velocity can be successfully captured and, consequently, the scaling behavior of the capture rate changes. In addition, for dark matter capture in the Earth (upper plots in \cref{fig:tFormEvaps}) the critical mass condition for collapse changes at $m_\chi \lesssim \SI{e12}{GeV}$ from being determined by $M_\text{sg}$ (requirement of sufficient self-gravity to overcome thermal pressure) to being given by $M_\text{f}$ (requirement of sufficient mass to overcome degeneracy pressure). This transition from $M_\textrm{f}$ to $M_\textrm{sg}$ is also the cause of the uneven spacing of the red evaporation contours for the Earth. Once $m_\chi$ is so low that \emph{all} dark matter is captured, the contours turn sharply upwards and become independent of $\sigma$, explaining the left edge of the wedges in \cref{fig:tFormEvaps}.

Besides this general behavior, we observe a number of interesting features in the gray contours in \cref{fig:tFormEvaps}. First, note that the shape of the uppermost contours for capture in Sun is somewhat different from the rest of the contours. This is particularly visible in the bottom left panel of \cref{fig:tFormEvaps}. The reason is that here the process that limits the black hole formation rate is no longer dark matter capture, but thermalization.

Second, note the unusual structure that is visible in the upper panels of \cref{fig:tFormEvaps} (capture in the Earth) at the low-$m_\chi$ cutoffs of the grey contours. The reason for this structure is that the Earth is denser at its $^{56}$Fe-rich core than in its mantle. Hence, the point where all dark matter particles crossing the core can be captured is reached at larger $m_\chi$ and lower $\sigma$ than the point where this is the case for mantle-crossing trajectories. Effectively, the grey contours can be viewed as a superposition of two contours: one for capture by the core, one for capture by the mantle. This feature is particularly pronounced in the case of isotope-dependent cross-sections (top left panel), as in this case the capture rate in the core benefits both from a higher density \textit{and} from the $A^4$ enhancement due to the presence of heavier elements in the core.

In general, we observe that black hole formation proceeds faster in the case of isotope-dependent cross-sections than for isotope-independent cross-sections due to the enhanced capture rate on heavy nuclei. To see this for the Earth, note that the ranges of the vertical axes in the top left and top right panels of \cref{fig:tFormEvaps} are very different. For the Sun, we observe an appreciable effect even though the only relevant heavy isotope there is \iso{He}{4} with a mass fraction of about 25\%.

Finally, we note that, upon comparing the black hole evaporation and formation contours in Figure \ref{fig:tFormEvaps}, it is evident that evaporation is always quicker than formation. Therefore, for essentially all parameter space, we need not consider the implications of forming more than one black hole at a time in the Sun or Earth. One possible exception to this is an unobservably small sliver of parameter space just to the right of the line labeled ``Stable BHs'', which marks where the dark matter accretion matches the black hole evaporation rate. In this small region, in principle it is also possible to form a black hole that neither grows nor evaporates away.

\section{Neutrinos from Evaporating Black Holes}
\label{sec:neutrino}

When a population of dark matter particles collapses into a black hole at the center of the Sun or the Earth and the black hole subsequently evaporates, the Hawking radiation emitted in the process offers an interesting target for detection. The main experimental signatures are anomalous heating of the Earth, which we will discuss in \cref{sec:results}, and neutrinos, which is the topic of this section. As both the formation and the evaporation of the black hole take place deep inside the Sun or Earth, the other components of the Hawking radiation cannot reach a detector. To the best of our knowledge, this is the first time neutrinos are discussed as a signature of black holes formed from dark matter.  The neutrino energy spectrum is an almost perfect thermal spectrum peaked at the temperature of the black hole, that is \cite{Hawking:1974rv, Page:1976df},
\begin{align}
    \tbh = \frac{1}{8 \pi G \mbh(t)} \,,
\end{align}
where we emphasize that the black hole mass $\mbh(t)$ decreases with time $t$. Hence, the neutrino radiation starts out at a low, often undetectable, temperature, but once the black hole mass drops below $\sim \SI{e-23}{M_\odot}$, it becomes hot enough ($\gtrsim \SI{500}{GeV}$) to be of interest to neutrino telescopes like IceCube. We do not consider lower energy neutrinos because of the overwhelming background of atmospheric neutrinos. Neglecting black hole accretion for the moment, the expression given in \cref{eq:bh3} for the black hole lifetime shows that the neutrino flux is detectable for the last $\sim \SI{4}{days}$ of the black hole's life. We will see below that an actual experimental search will have far better reach using the high-energy tail of the Hawking radiation spectrum.

The expected neutrino spectrum has the following characteristics:
\begin{enumerate}
    \item The \textbf{spectral shape} is close to a black-body spectrum, superimposed with a spectrum of secondary neutrinos from the decays of heavier components of the Hawking radiation.  The signal thus  has a very different energy dependence than the atmospheric neutrino background, which is a falling power law.
    
    \item Obviously, the signal is \textbf{strongly directional}, originating from the center of the Sun. At the high neutrino energies that we are most interested in, the correlation between the a priori unknown primary neutrino direction and the observed direction of the secondary lepton is very strong, and moreover the angular resolution of the detector is quite good. Directionality is therefore crucial in discriminating against the essentially isotropic atmospheric neutrino background.
    
    Note that Directionality is not useful to suppress the background of so-called ``solar atmospheric neutrinos'', that is neutrinos produced by cosmic ray interactions in the outer layers of the Sun~\cite{Seckel:1991ffa, Moskalenko:1993ke, Ingelman:1996mj, Hettlage:1999zr, Fogli:2006jk, Ng:2017aur, Arguelles:2017eao, Aartsen:2019avh}. However, it turns out that the solar atmospheric background exceeds the terrestrial atmospheric background only at neutrino energies in excess of several TeV. We will see that neutrinos with such high-energies are not of interest to us because, after being produced in the core of the Sun, they are efficiently absorbed on their way out and would therefore not reach a detector on Earth. For this reason, we will neglect solar atmospheric neutrinos in what follows.  They are in particular irrelevant for the limits we will derive based on IceCube data from Ref.~\cite{Aartsen:2016zhm} as the data in this reference is only presented in energy-integrated form and is therefore entirely dominated by low-energy terrestrial atmospheric neutrinos.
    
    \item The signal is \textbf{transient}: we expect neutrinos from black hole evaporation to come in bursts whose duration is much shorter than the typical duration of an observation run in a neutrino telescope. We will, however, find below that the transient nature of the black hole evaporation signal is a useful tool only in a limited region of parameter space, spanning about an order of magnitude in dark matter mass around $m_\chi \sim \SI{e17}{GeV}$. For other dark matter masses, black hole formation and evaporation is either too rare to be observable, or the sensitivity is so good that a time-integrated analysis is sufficient, or black holes evaporate so frequently that the neutrino is quasi-continuous.
    
    \item The signal is \textbf{flavor-universal}, as Hawking radiation populates all available degrees of freedom equally.  We do not expect this feature to be particularly relevant or useful, though, because IceCube's sensitivity to the flavor composition of the high-energy neutrino flux is rather poor.
\end{enumerate}

In what follows we describe the ingredients and assumptions made in determining the total event rate in IceCube:
\begin{description}[leftmargin=0cm,style=unboxed]
    \item[Instantaneous primary neutrino spectrum]
    We begin by considering the instantaneous primary neutrino flux from an evaporating black hole, which is given by
    \begin{align}
        \frac{\diff^3 N_\alpha^\text{prim}}{\diff E \,\diff t\, \diff A}
            &= \frac{1}{4\pi r^2} \frac{\Gamma_{s=1/2}(\mbh,E)}{2\pi}
               \frac{g_\alpha}{e^{E/\tbh}+1}\,.
        \label{eq:nu-flux}
    \end{align}
    Here, $\diff N_\alpha/(\diff E\,\diff t\,\diff A)$ denotes the flux of neutrinos of flavor $\alpha$, $E$ is the neutrino energy, $r$ is the Earth--Sun distance, and $g_\alpha = 2$ denotes the number of internal degrees of freedom (left-handed neutrino and right-handed anti-neutrino). Of course, if neutrinos are Dirac particles, also right-handed neutrino and left-handed anti-neutrino states would be produced by evaporating black holes, but as these are undetectable, we ignore them here. We also do not distinguish between neutrinos and anti-neutrinos as IceCube does not have this capability (except for weak discrimination power on a statistical basis, see Ref.~\cite{Ribordy:2013xea}).  The dimensionless factor $\Gamma_{s_\alpha}$ in \cref{eq:nu-flux} is once again the grey-body correction introduced in \cref{eq:grey-body} and evaluated in \texttt{BlackHawk}~\cite{Arbey:2019mbc}.

    \begin{figure}
        \centering
        \includegraphics[width=0.67\linewidth]{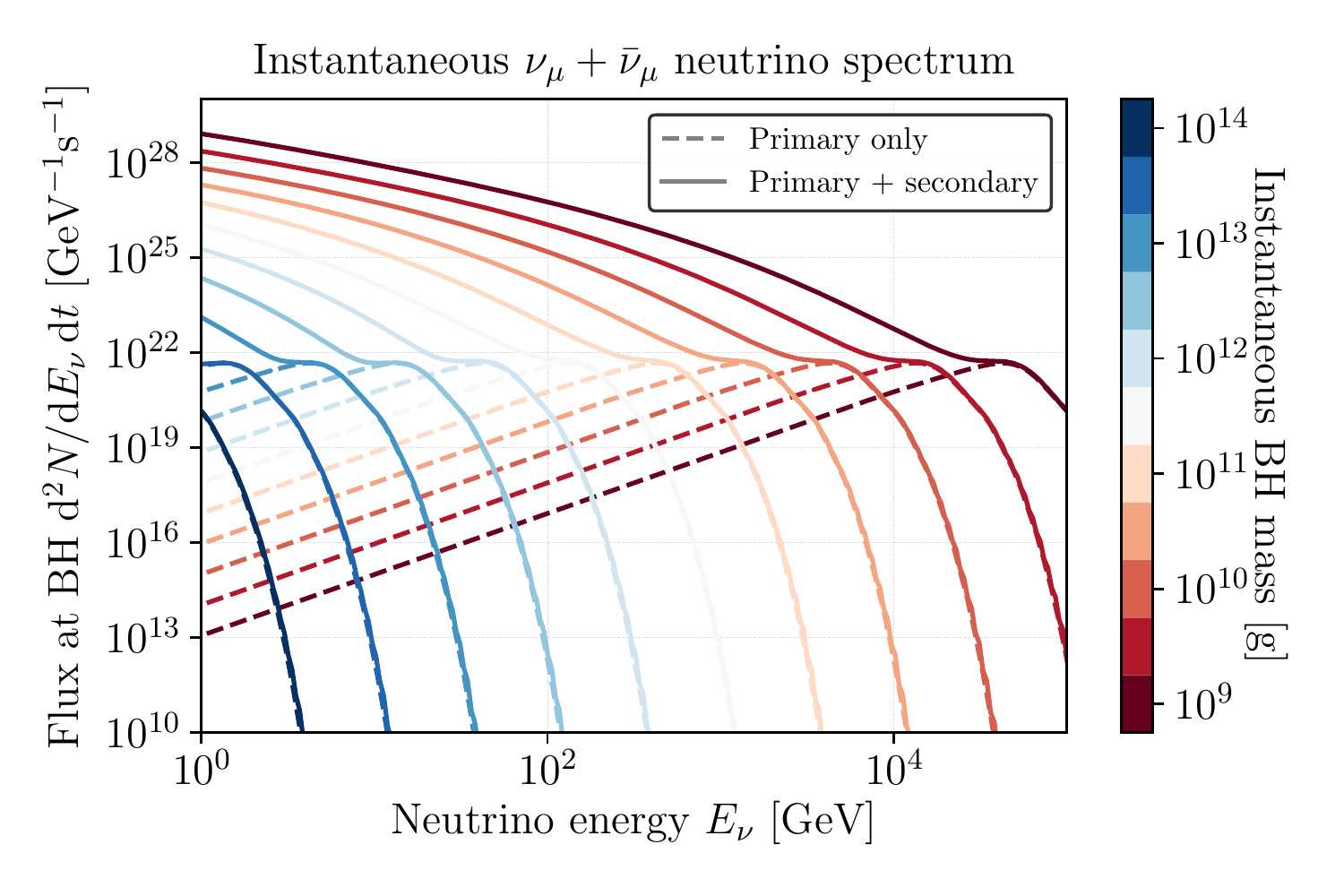}
        \caption{Instantaneous muon-neutrino spectrum for black holes of different masses, as indicated by the color code. The dashed lines contain only the primary contribution from the Hawking radiation while the solid lines contain both the primary and secondary spectrum of neutrinos, where the secondaries arise from decays and hadronization of heavier non-neutrino states in the primary spectrum.}
        \label{fig:inst_BH_neutrino_spectra1}
    \end{figure}

    \item[Instantaneous secondary neutrino spectrum]
    Many of the particles generated at the event horizon through Hawking radiation are unstable, and their decays produce a secondary flux of neutrinos. For instance, high-energy quarks will undergo parton showering and hadronization, followed by the decays of unstable hadrons.  Secondary neutrinos typically dominate the low energy part of the neutrino spectrum. Their flux is given by
    \begin{align}
      \frac{\diff^3 N_\alpha^\text{sec}}{\diff E \,\diff t\, \diff A}
          &= \int \! \diff E^\prime \,
             \sum_{i} \frac{\diff^3 N_i^\text{prim}}{\diff E^\prime \,\diff t\, \diff A}
                      \frac{\diff N_\alpha^i}{\diff E} \,,
    \end{align}
    where the sum runs over all primary particles whose decay or hadronization leads to the production of a neutrino state $\nu_\alpha$.  As before, $\diff^3 N_i^\text{prim} / (\diff E^\prime \,\diff t\, \diff A)$ are the primary particle fluxes, and the coefficients $\diff N_\alpha^i/\diff E$ describe the contribution to the $\nu_\alpha$ flux from primary particles of type $i$. These coefficients are implemented in \texttt{BlackHawk} through the inclusion of decay and hadronization tables, which have been generated in \texttt{PYTHIA} \cite{Sjostrand:2014zea}. The details of these table can be found in Appendix~B.3 of Ref.~\cite{Arbey:2019mbc}. The key limitations of this procedure are the limited energy domain of the hadronization tables, $[0.1,10^5]~\SI{}{\GeV}$, and the missing electroweak effects for highly energetic primary particles.  We mitigate this problem by extrapolation, using the fact that, far away from any kinematic thresholds, the ratio of the total neutrino flux (primary plus secondary) to the primary-only flux scales as a simple power law when plotted as a function of neutrino energy times black hole mass (i.e.\ $E_\nu \mbh$).  This allows for an estimate of the neutrino flux for regions where the majority of the black hole's lifetime is spent at temperatures above \SI{E+5}{\GeV}, that is for $\mbh^\text{init} \lesssim \SI{E+8}{\gram}$.
    
        \begin{figure}
        \centering
        \includegraphics[width=0.67\linewidth]{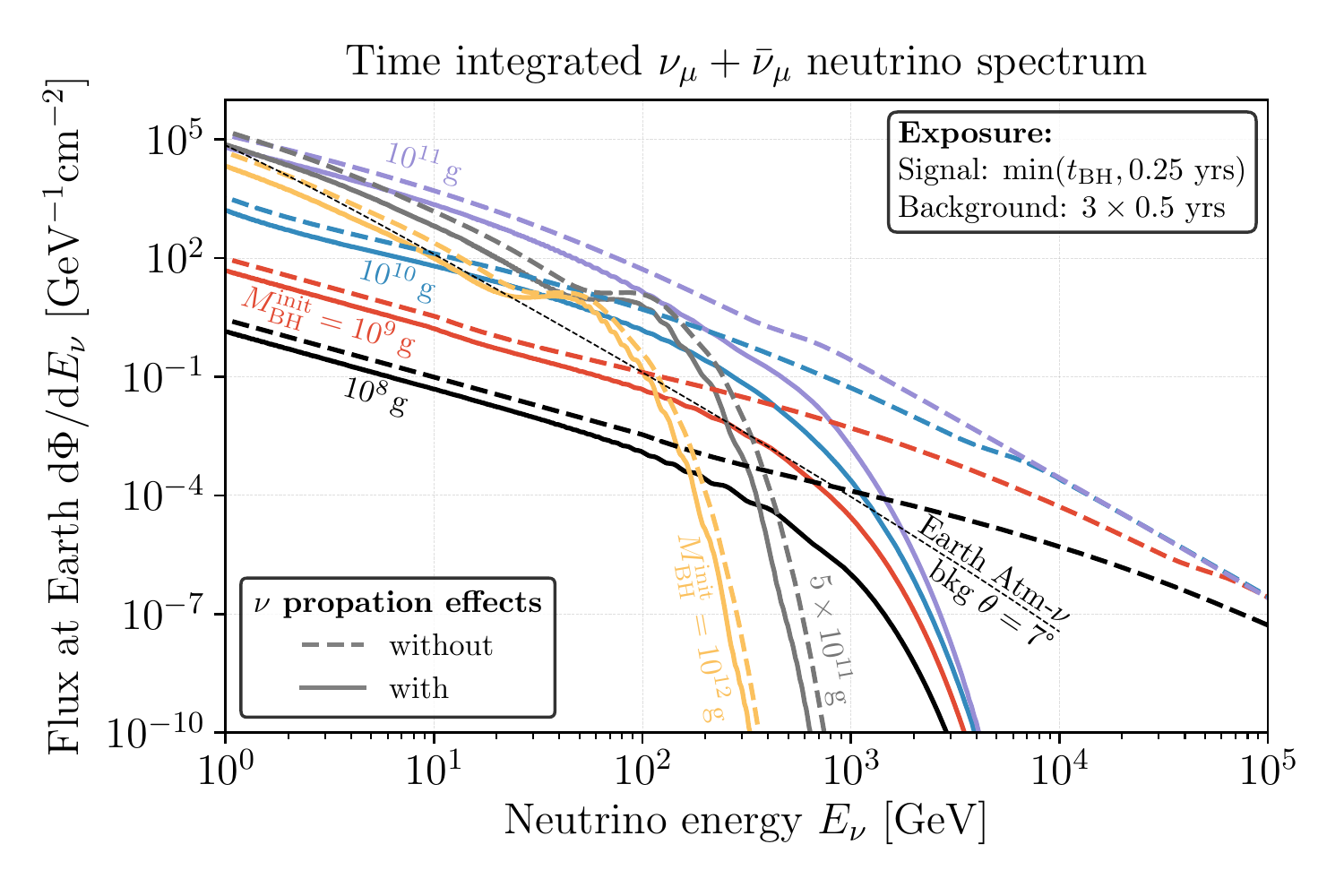}
        \caption{Flux of muon-neutrinos at Earth, time-integrated over the minimum of either the entire lifetime of the black hole or the first \SI{0.25}{\years} of its life, i.e.\ $\min(\timebh, \SI{0.25}{\years})$. This is a conservative estimate of the neutrino flux from a single black hole contributing to the IceCube analysis in \cite{Aartsen:2016zhm}. Solid (dashed) lines are with (without) the effects of neutrino propagation through the Sun and oscillation in vacuum (see text for details). }
        \label{fig:tot_BH_neutrino_spectra1}
    \end{figure}

    In \cref{fig:inst_BH_neutrino_spectra1} we show the instantaneous primary and secondary neutrino spectra for different black hole masses.  We focus here on the muon neutrino flux only as this is the flavor that is most easily detectable in neutrino telescopes like IceCube and SuperKamiokande. As expected the temperature of the black hole, inversely proportional to its mass, sets the energy at which the primary spectrum is peaked. The secondaries serve to increase the flux by several orders of magnitude at energies below this peak, largely due to decays of other energetic species. A key observation is that for black hole masses below \SI{E+9}{\g} the spectrum is peaked at energies in excess of \SI{E+5}{\GeV}. This is the region where \texttt{BlackHawk} is no longer able to make fully reliable predictions for the secondary spectrum of neutrinos as the dominant component of the primary flux is outside of the range of the hard-coded decay and hadronization tables. Instead, the conservative extrapolation procedure described above needs to be invoked.
    
    \item[Time-integrated neutrino spectrum]
    To compare to time-integrated experimental data, we need to track the mass-evolution of the black hole and integrate the neutrino spectrum over time.  In doing so, we will assume that accretion is subdominant compared to the Hawking evaporation rate, that is we will neglect the first two terms on the right-hand side of \cref{eq:bhtotal}. In other words, we will describe the time-evolution of the black hole mass using the differential equation \cite{MacGibbon:1990zk,Arbey:2019mbc}
    \begin{align}
        \frac{\diff\mbh}{\diff t} = -\frac{f(\mbh)}{(G\mbh)^2} \,,
        \label{eq:dMdt-page}
    \end{align}
    with the Page factor $f(\mbh)$ from \cref{eq:page-factor} taking into account the number of degrees of freedom in the Hawking radiation as well as grey-body corrections.  We integrate \cref{eq:dMdt-page} over the entire black hole evolution from its initial mass $\mbh^\text{init}$ until the point where $\tbh = \SI{E+5}{\GeV}$, beyond which we make use of the extrapolation for the secondary spectrum detailed above.  At each time-step in the black hole evolution, we calculate the instantaneous neutrino spectrum. This allows for a simple integration over time to determine the total neutrino spectrum. For the analysis performed here, we integrate the neutrino flux over a time interval corresponding to either the black hole lifetime, or \SI{0.25}{\years}, whichever is shorter. This limit of \SI{0.25}{\years} is largely for illustration purposes because, as we will see later, such long-lived black holes are not able to evaporate in the Sun. The reason for this is that the dark matter accretion rate required to form at least one black hole over the duration of the IceCube data taking period would imply that the accretion rate onto the black hole is larger than its evaporation rate.
    
    \cref{fig:tot_BH_neutrino_spectra1} shows the resulting time-integrated neutrino spectra on Earth for different values of the initial black hole mass as well as the expected background of atmospheric neutrinos using the South Pole winter results from Ref.~\cite{Honda:2015fha} (assuming a fixed energy independent angular resolution of \SI{7}{\degree}). Let us for the moment focus on the dashed lines in the figure, which neglect the effects of neutrino propagation in the Sun and later in vacuum. There are two key features of these spectra. Firstly, for initial black hole masses where the evaporation happens in less than \SI{0.25}{\years}, i.e.\ $\mbh \lesssim \SI{E+11}{\g}$, we observe the final high-energy tail of the neutrino spectrum. In this regime, for decreasing values of the initial black hole mass the amplitude of the spectrum decreases largely due to the smaller lifetime of the black hole. However, at sufficiently high neutrino energies there is convergence of the spectra even for different initial black hole masses. This follows as the high-energy tail of the spectrum is mostly populated during the very final stages of evaporation, which leads to identical spectra in this region for all black holes that completely evaporate within the observation period. For black holes that are longer lived than \SI{0.25}{\years}, the resulting neutrino spectrum is radically different. For the cases depicted, $\mbh^\text{init} = \{\SI{5E+11}{},\SI{E+12}{}\}\,\si{\g}$, the black hole mass does not evolve appreciably over the \SI{0.25}{\years} time interval. The spectrum therefore bears resemblance to the instantaneous spectra from the left-hand panel, including the absence of the high-energy tail. 
    
    \begin{figure}
        \centering
        \includegraphics[width=0.67\linewidth]{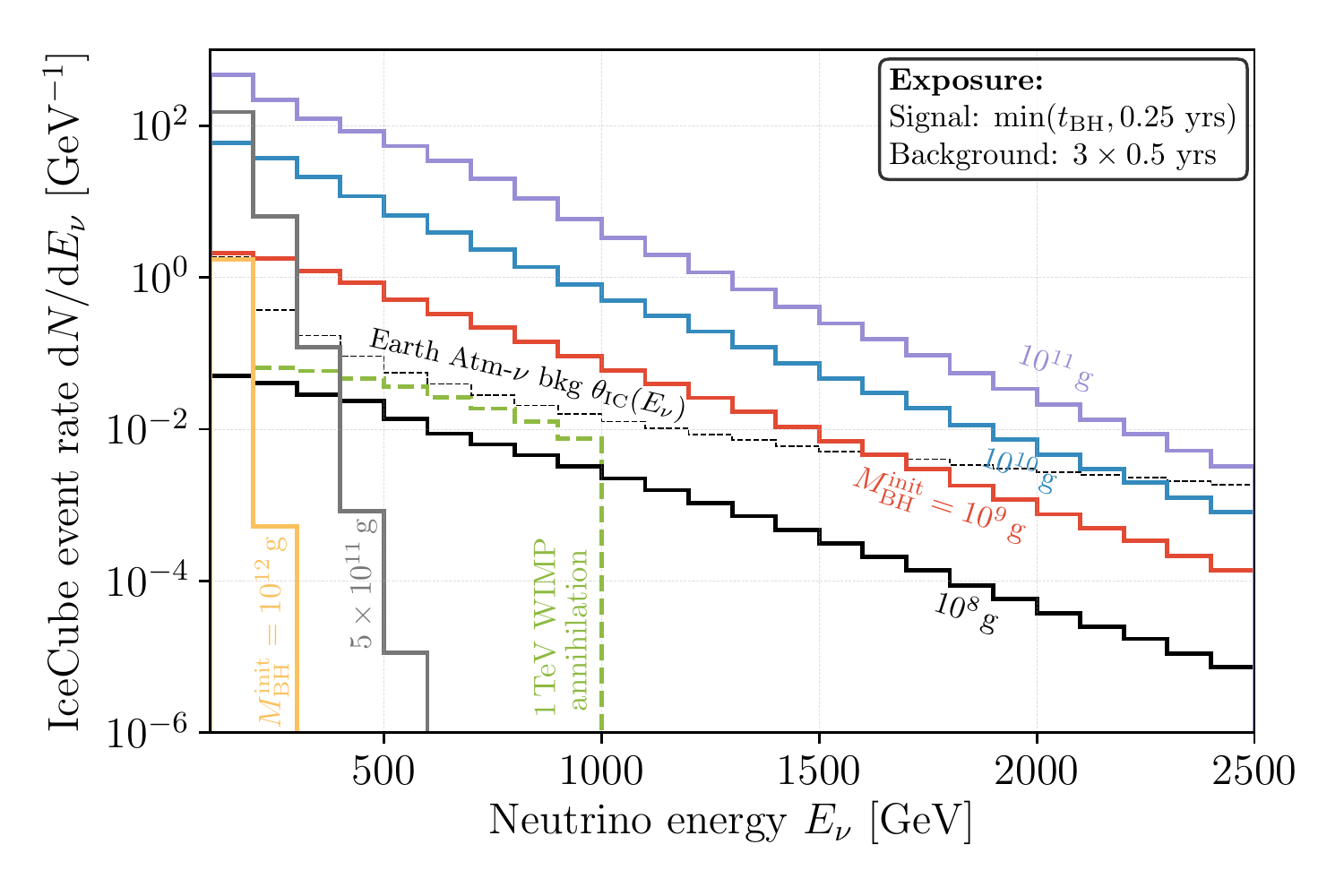}
        \caption{Differential event rate in IceCube as a function of the muon neutrino energy. For comparison we also show the event rate for the case of a \SI{1}{\TeV} WIMP annihilating to $W^+W^-$ in the Sun from Ref.~\cite{Aartsen:2016zhm}.}
        \label{fig:IC_energy_rate}
    \end{figure}
    
    \item[Solar neutrino propagation] With the primary and secondary neutrino spectra at hand, the final step to determine the flux on Earth is the propagation of the neutrinos from the center of the Sun. There are three main effects that must be captured to describe the propagation accurately: ($i$) neutrino trapping through scattering off the stellar medium, ($ii$) neutrino oscillations, and ($iii$) tau neutrino regeneration, that is the production of tertiary neutrinos from the decays of $\tau$ leptons created in $\nu_\tau$ interactions inside the Sun. The effects of neutrino trapping can be approximated through the use of an exponential cut-off depending on the number of scattering targets and the energy dependent scattering cross-section of the specific neutrino species in question. The effects of tau neutrino regeneration, however, soften this exponential cut-off (see for example Ref.~\cite{Albuquerque:2000rk}), i.e.\ production and decay of tau leptons shift the neutrino spectrum to lower energies, allowing for a larger fraction to escape from the Sun. To better model these effects as well as to include neutrino oscillations in matter, we have utilized the public code \texttt{nuSQuIDS}~\cite{Delgado:2014kpa, squids, nusquids}. The time integrated neutrino spectrum for each flavor of neutrino is used as an input, whereby \texttt{nuSQuIDS} propagates the flux utilizing the standard solar model BS05(AGS,OP) of Ref.~\cite{Bahcall:2004pz}. The results of this procedure are shown as solid lines in \cref{fig:tot_BH_neutrino_spectra1}. The key feature is of course the opacity of the Sun to high-energy neutrinos due to which the flux of TeV energy neutrinos is strongly suppressed. This highlights the importance of the secondary neutrino flux that dominates at sub-TeV energies. Secondly, we see a mismatch between the curves with and without solar propagation at lower neutrino energies, which arises from neutrino oscillation effects. More precisely, as the initial flux of tau flavored neutrinos (not shown here) is subdominant compared to electron and muon flavors, neutrino oscillations lead to a net depletion of the $\nu_\mu$ flux and thus to the observed mismatch.

    \begin{figure}
        \centering
        \includegraphics[width=0.67\linewidth]{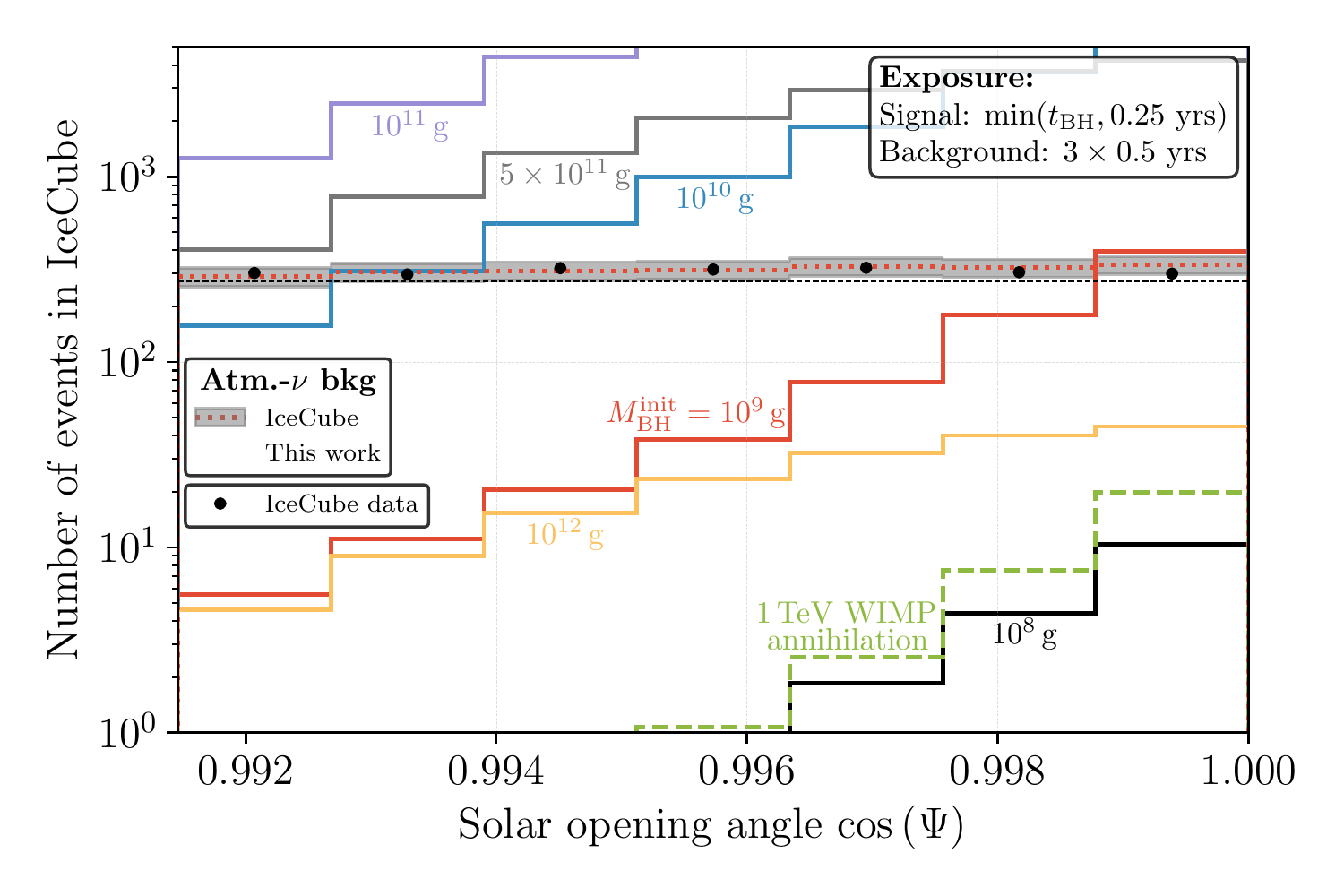}
        \caption{Event rate as a function of the cosine of the opening angle towards the Sun for the same spectra as in \cref{fig:IC_energy_rate}. The black dots are the observed event numbers in three years of IceCube data from Ref.~\cite{Aartsen:2016zhm}, while the grey shaded band is the expected background from atmospheric neutrinos including its $2\sigma$ statistical uncertainty.}
        \label{fig:IC_ang_rate}
    \end{figure}

    \item[IceCube event rate]
    To place limits on the final muon neutrino flux, we have recast an IceCube search for neutrinos from dark matter annihilation, presented in Ref.~\cite{Aartsen:2016zhm}. Using the effective area given in Fig.~4 of that reference, the spectra from our \cref{fig:tot_BH_neutrino_spectra1} can be used to determine a differential event rate with respect to truth level neutrino energy.
    We restrict our analysis to detection at IceCube rather than lower-threshold detectors like Super-K \cite{Abe:2020sbr} because, as shown in \cref{fig:tot_BH_neutrino_spectra1}, atmospheric neutrino backgrounds dominate at lower energies, and thus we do not expect a significant improvement over the results presented here.
    This event rate is shown in \cref{fig:IC_energy_rate} for the relevant energy domain, and compared to the predicted neutrino spectrum from WIMP annihilation to $W^+W^-$, the example used as a benchmark in Ref.~\cite{Aartsen:2016zhm}. While IceCube themselves have utilized an unbinned likelihood analysis to set limits on this benchmark model, they only present their data binned as a function of the angle $\Psi$ between the reconstructed neutrino arrival direction and the location of the Sun in the sky. To determine the differential event rate with respect to $\Psi$ for neutrinos from black hole evaporation, we use IceCube's energy-dependent median angular resolution (right-hand panel of Fig.~4 in Ref.~\cite{Aartsen:2016zhm}) to build Gaussian distributions centred around zero, i.e.\ the exact position of the Sun, for every value of the truth level neutrino energy. We have verified that, using the WIMP annihilation spectrum, this method faithfully reproduces the resulting angular distribution shown in Fig.~6 of Ref.~\cite{Aartsen:2016zhm}.  
 
    \begin{figure}
        \centering
        \includegraphics[width=0.67\linewidth]{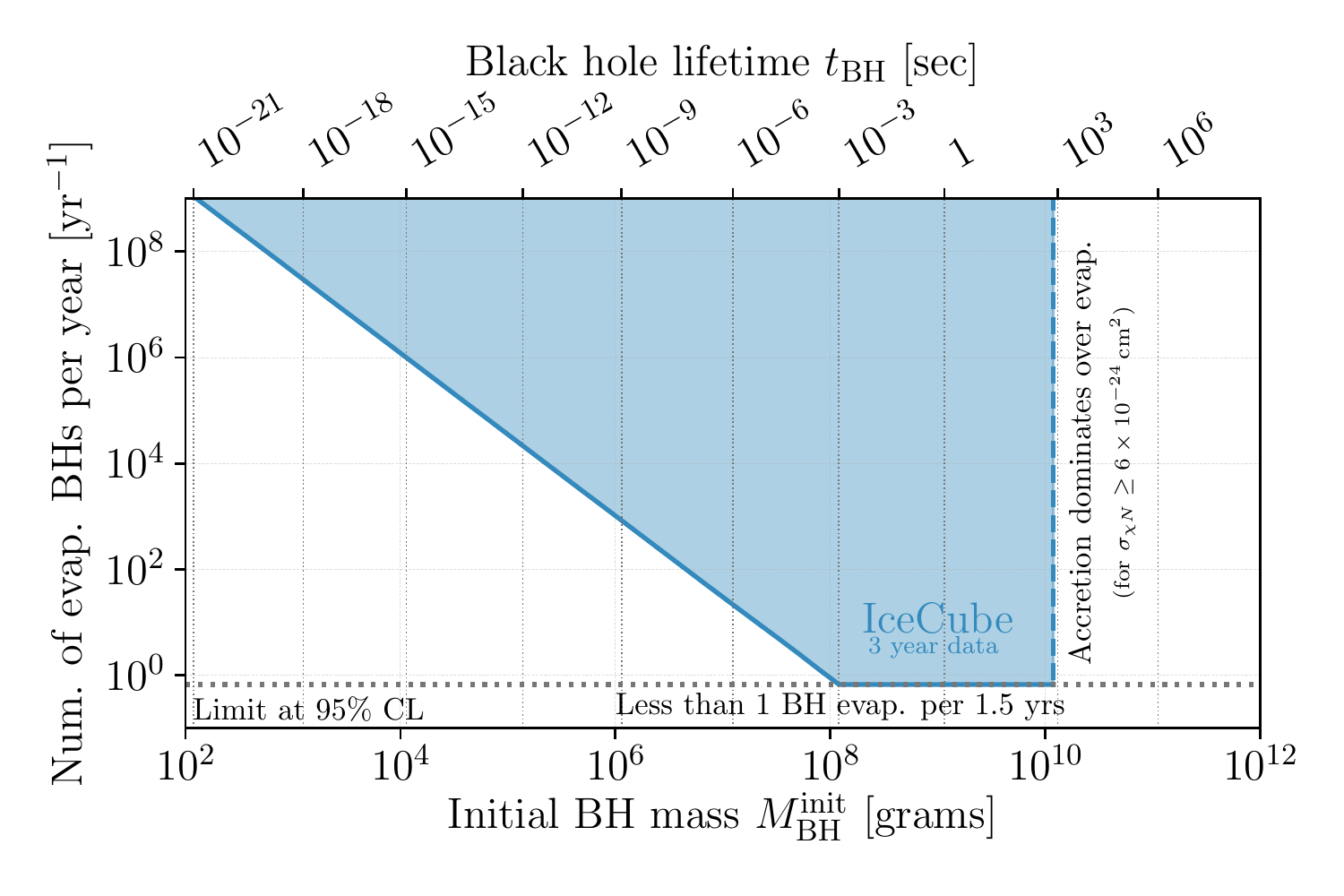}
        \caption{Excluded values (blue shaded region) of initial mass black holes formed in the Sun as a function of the total number that evaporate over the \SI{1.5}{\years} of IceCube exposure. This is based on a recast of the IceCube analysis from Ref.~\cite{Aartsen:2016zhm}, see text for details.}
        \label{fig:limit_num_BHs_evap}
    \end{figure}
 
    The resulting number of IceCube events as a function of this angular variable is shown in \cref{fig:IC_ang_rate}. We also overlay both the IceCube data (black points) as well as the atmospheric neutrino background prediction (the red-dotted line is the central value, while the grey band shows the $2\sigma$ statistical uncertainties). As expected, harder neutrino spectra lead to distributions that are more sharply peaked at $\cos\Psi = 1$, while the softer spectra for heavy initial mass black holes lead to flatter distributions. Performing a binned likelihood analysis, all illustrated spectra are excluded at 95\%~CL except for the $\mbh^\text{init}=\SI{E+8}{\gram}$ (black) spectrum. For such small values of the initial black hole mass, however, we typically expect a number of these black holes to both form and completely evaporate within the exposure time of the IceCube data taking (see \cref{fig:tFormEvaps} above). We therefore also determine limits for lighter initial mass black holes as a function of the number of evaporated black holes with a given initial formation mass. The blue shaded region in \cref{fig:limit_num_BHs_evap} violates the derived limit. In addition shown in grey are contours of constant black hole lifetime for a given initial black hole mass. We see that between approximately $\mbh^\text{init} \in [10^8,\SI{4E+10}{}]~\SI{}{\gram}$ a single black hole evaporating is excluded by IceCube, while for lighter initial black hole masses a larger number must evaporate over the \SI{1.5}{\year} exposure period to be detectable. We do not exclude black holes with $\mbh^\text{init} > \SI{4E+10}{\gram}$ (vertical dashed blue line). These heavier black holes cannot evaporate in the Sun as the dark matter accretion rate onto the black hole is too large for DM--nucleon cross-sections $\sigma_{\chi N} \geq \SI{7E-24}{\cm^2}$. This cross-section corresponds to the minimum cross-section such that at least one black hole per year is formed in the Sun.
\end{description}

\section{Results and Discussion}
\label{sec:results}

\begin{figure}
    \centering
    \includegraphics[width=\linewidth]{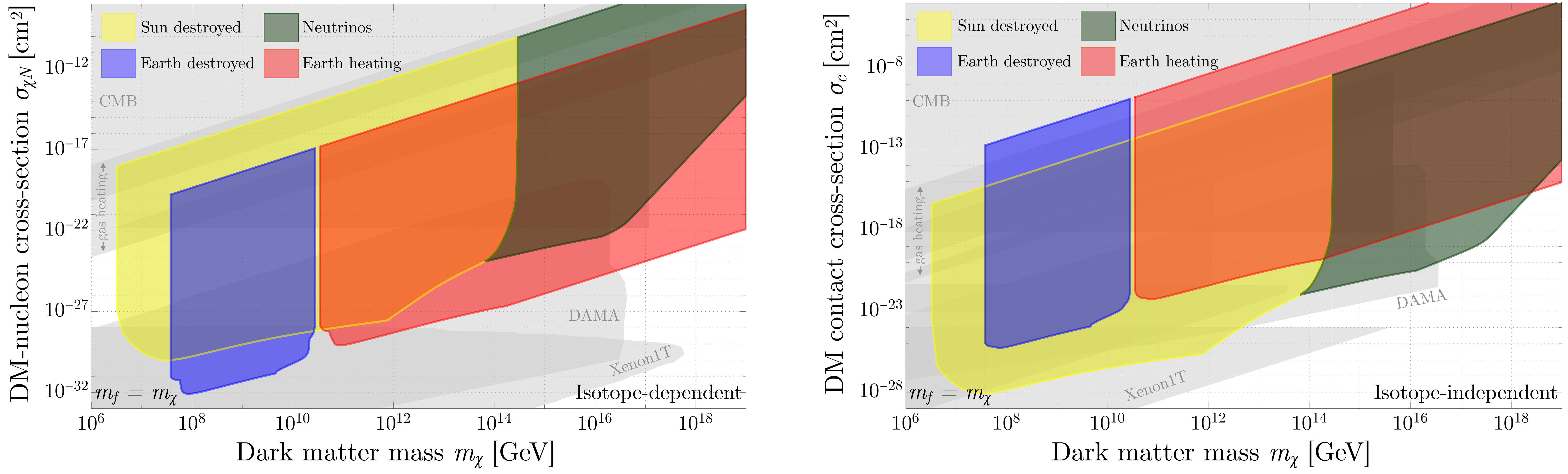}
    \caption{
    {\bf (left)} Solar and terrestrial bounds on the DM--nucleon scattering cross-section for non-annihilating dark matter with conventional spin-independent (and therefore isotope-dependent) couplings. The blue and yellow regions are excluded by the formation of a black hole in the Earth or Sun that would grow and consume these bodies within a billion years. The red region is excluded by the formation of evaporating black holes in the Earth that would result in more than the observed \SI{44}{TW} of heat emanating from the Earth's surface. Finally, the dark green region is excluded by the null observation of a high-energy flux of neutrinos that would be produced by black holes evaporating in the Sun. The edges of the exclusion regions can be understood as follows: for $m_\chi \lesssim \SI{e7}{GeV}$, black holes will not form, given the amount of dark matter collected in a Gyr. The upper cutoff in $m_\chi$ is given by the Planck scale, $\SI{e19}{GeV}$. The lower edges of the exclusion regions are determined by requiring that dark matter collects, cools, and collapses to a black hole in a Gyr. Their upper edges are determined by requiring that dark matter can drift to the center of the Earth or Sun against the viscous drag of nuclei in less than a Gyr. Previous limits from underground direct detection experiments are shaded in gray~\cite{Bernabei:1999ui,Clark:2020mna,Cappiello:2020lbk,Kavanagh:2017cru}, as are CMB bounds \cite{Gluscevic:2017ywp, Dvorkin:2013cea}, bounds from the heating of interstellar gas clouds \cite{Bhoonah:2018gjb,Bhoonah:2020dzs}, and bounds from searches for DM tracks in ancient mica minerals~\cite{Bramante:2018tos,Price:1986ky}. {\bf (right)} Same as left, but for a dark matter--nucleus scattering via isotope-independent contact interactions, as discussed around \cref{eq:losscomp}. We assume here that, even though dark matter may consist of composite objects, the constituent masses are large enough for Pauli blocking to be irrelevant during collapse.}
    \label{fig:dmnucleon}
\end{figure}

Having discussed the evolution of dark matter captured in the Sun or Earth, the dynamics of black hole formation and evaporation, and the constraints arising from the neutrino component of the Hawking radiation, we are now ready to put all of these results together. We do so in \cref{fig:dmnucleon}, where we show bounds on the DM--nucleon scattering cross section in the case of conventional isotope-dependent DM--nucleus scattering (left panel), and on the universal dark matter scattering cross section in models with isotope-independent couplings such as certain asymmetric composite dark matter models (right panel).

We first discuss the constraints arising from the fact that the Sun and Earth have not yet been consumed by a black hole (yellow and blue shaded regions in \cref{fig:dmnucleon}). The parameter region excluded by these bounds satisfies the following criteria:
\begin{enumerate}
    \item At least one black hole must have formed within the age of the solar system, $\sim \mathcal{O}(\si{Gyr})$. This means that dark matter capture, thermalization, and collapse must terminate after at most $\sim \SI{1}{Gyr}$. In most of the parameter space, the limiting process is capture, which therefore defines the lower and left edges of the yellow and blue exclusion regions in \cref{fig:dmnucleon}, as can be read off directly from \cref{fig:tFormEvaps}.
    
    \item The DM--nucleus scattering cross section must be small enough for dark matter thermalization to be efficient. As discussed in \cref{sec:drifttime} -- see in particular \cref{eq:sigmaDrift} -- if the time it takes the dark matter to drift to the center of the Sun or Earth exceed $\sim \SI{1}{Gyr}$, a black hole will never form. This condition defines the upper boundary of the yellow and blue exclusion regions in \cref{fig:dmnucleon}.
    
    \item The black hole evaporation time should be longer than the age of the solar system, $\sim \si{Gyr}$.  This condition sets the right boundary of the shaded yellow and blue exclusion regions in \cref{fig:dmnucleon}. Note that there is a tiny sliver of parameter space between the blue and red shaded regions in which a black hole will form, but accretion will be too inefficient for it to consume the Earth within \SI{1}{Gyr}. We ignore this small region here.
\end{enumerate}
It is gratifying to note that the dark matter--nucleus interaction parameter space for which a small black hole would be growing inside the Earth at present can be ruled out by underground searches for dark matter and the continued survival of the Sun.

More benign black holes can be formed by dark matter with masses larger than \SI{e11}{GeV} in the Earth or \SI{e15}{GeV} in the Sun. These benign black holes are small enough that they evaporate faster than they accrete their host's material.  Nevertheless, part of their parameter space can be excluded because the energy released by their evaporation would significantly increase the heat flow emanating from the Earth. As noted in Refs. \cite{Mack:2007xj,Bramante:2019fhi}, the maximum heat flux observed in the Earth is about \SI{44}{TW} \cite{williams1974, lister1990, JAUPART2015223, Davies1980, sclater1980, pollack1993, davies2010} (although some works find a lower value, see e.g.\ Ref.~\cite{arevalo2009k}). While much of this heat flux has been attributed to radiogenic elements in the Earth such as $^{235}$U, we conservatively exclude only dark matter models which would form evaporating black holes that would release $\gtrsim \SI{44}{TW}$ of heat into the Earth. In practice, this requires a capture rate $C_\chi \gtrsim \SI{44}{TW} / m_\chi$. We furthermore impose the restriction that these black holes must form and evaporate within $\sim1$ kyr. These timescales are short enough that, due to the Earth's heat capacity and the fact that it radiates heat as a black body, we expect the observed heat flux today would be steady \cite{arevalo2009k}.

The dark green region in \cref{fig:dmnucleon} represents the IceCube constraint on the neutrino flux from black hole evaporation derived in \cref{sec:neutrino}. 
To obtain this curve, we determined for each parameter point $(m_\chi, \sigma_{\chi N})$ the number of black holes that form within the \SI{0.25}{\years} of IceCube data taking based on the formation timescales given in \cref{fig:tFormEvaps}. We then calculated the initial mass of these black holes, based on \cref{eq:Mcrit}, and read off from \cref{fig:limit_num_BHs_evap} whether or not that parameter point's corresponding black hole evaporation time allows it to be excluded.

In the right-hand panel of \cref{fig:dmnucleon}, we show the analogous bounds for an isotope-independent DM--nucleus contact cross-section, as would be appropriate for large composite dark matter, see \cref{eq:losscomp} and the surrounding discussion. In this case, there is no coherent enhancement for scattering on large nuclei. As a consequence, for a fixed DM--nucleus scattering cross-section, the solar bounds on black hole formation are more restrictive since for a fixed nuclear scattering cross-section, the Sun is more massive and has a greater stopping power than the Earth. Note that in \cref{fig:dmnucleon} (right), we assume that the constituent masses of the composite dark matter objects are large, so that the critical mass for collapse, \cref{eq:chandraMass}, is given by the self-gravitating mass $M_\text{sg}$, not by the maximal mass $M_f$ at which Fermi degeneracy pressure can stabilize the dark matter core and prevent collapse.

\begin{figure}
    \centering
    \includegraphics[width=\linewidth]{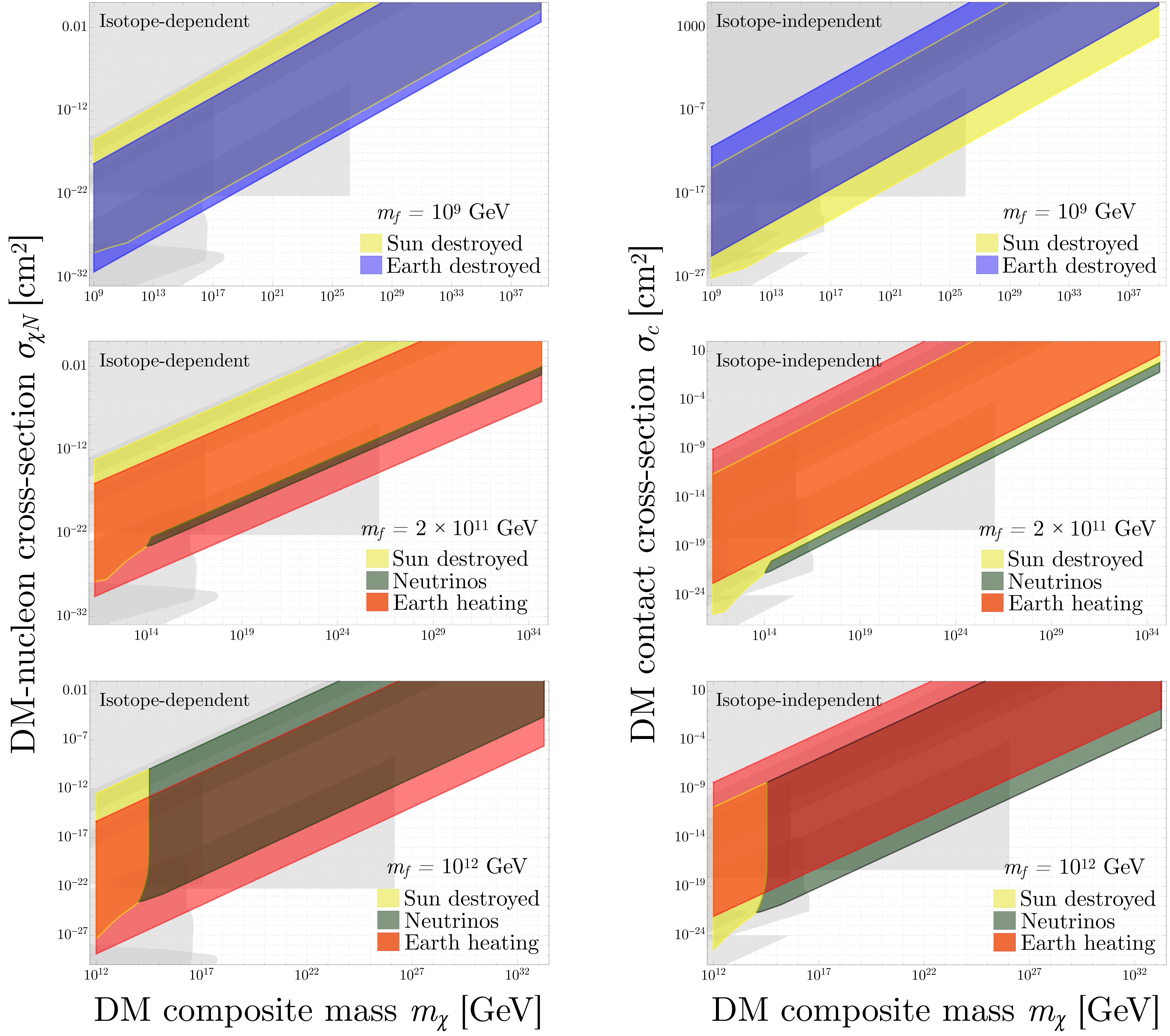}
    \caption{Same as Figure \ref{fig:dmnucleon}, except here we consider composite asymmetric dark matter with a mass $m_\chi$, composed of fermions with masses $m_f$, where $m_f$ fixes a minimum black hole mass through \cref{eq:chandraMass}. This allows us to extend the plots to much larger dark matter masses than in \cref{fig:dmnucleon}, which we cut off at the Planck scale. Panels on the left are for conventional spin-independent (and isotope-dependent) DM--nucleus scattering, while panels on the right are for isotope-independent scattering. For $m_f = \SI{e9}{GeV}$ (top row), all black holes formed are large enough to consume the Earth and Sun within a Gyr. For $m_f = \SI{2e11}{GeV}$ (middle row), black holes formed in the Earth and Sun will either evaporate or grow depending on how much additionally captured dark matter falls into the black hole. For $m_f = \SI{e12}{GeV}$ (bottom row), black holes formed in the Earth and Sun tend to evaporate.}
    \label{fig:comp}
\end{figure}

Next we relax this assumption and consider how solar and terrestrial black hole formation bounds change if we assume dark matter consists of composite objects of mass $m_\chi$, but composed of fermions with smaller fundamental mass $m_f$. Lowering this constituent mass $m_f$ increases the minimum mass of the black holes formed in the Sun and Earth since the dark matter's Chandrasekhar mass given by \cref{eq:chandraMass} then depends on $m_f$.

In \cref{fig:comp} we show bounds on composite dark matter with either conventional isotope-dependent couplings to nuclei (left panels) or with fixed contact DM--nuclear interaction (right).  We consider three illustrative values for the constituent mass, $m_f = 10^9$, $2\times10^{11}$, and  $\SI{e12}{GeV}$. For the case that $m_f = \SI{e9}{GeV}$, the minimum black hole mass is large enough that all black holes formed in the Earth and Sun will consume and destroy their host (top panels in \cref{fig:comp}). For $m_f = \SI{2e11}{GeV}$ (middle panels), whether the black holes formed evaporate or grow to consume their hosts depends on how quickly they accrete additional dark matter, which in turn depends on the composite--nuclear cross-section. For the case that $m_f = \SI{e12}{GeV}$, the black holes formed evaporate over most parameter space, leading to Earth heating and neutrino flux bounds extending to dark matter composite masses up to $m_{\chi} = \SI{e33}{GeV}$.

The high-$m_\chi$ cutoffs in these exclusions are set by the Chandrasekhar critical mass condition, this time applied to the composite itself. If $m_\chi\geq M_\textrm{pl}^3/m_f^2$, then the composites in question will collapse into black holes. In theory, there could also be a flux-limitation to such plots, wherein the total dark matter flux through the stellar body within the appropriate timescale would become less than $2m_\chi$, but for the chosen values of $m_f$, this is never the limiting factor.

\section{Conclusions}
\label{sec:conclusions}

To conclude, we have derived a suite of novel bounds on heavy ($> \SI{e7}{GeV}$), non-annihilating dark matter based on the fact that such dark matter particles, when captured by the Sun or Earth, could accumulate in the solar or terrestrial core in sufficient quantity to form black holes. These black holes can either grow by consuming their host, or they can evaporate. In the former case, the continued existence of the Sun and the Earth can be used to set limits on the dark matter mass and capture cross section. In the latter case, limits can be derived based on the observed heat flux from the Earth, and from the non-observation of neutrinos from black hole evaporation in IceCube. For the latter constraint, we have recast an existing IceCube search for dark matter annihilation in the Sun~\cite{Aartsen:2016zhm}, but we have also demonstrated that a dedicated search for evaporating black holes can improve the sensitivity significantly.

Our main results are summarized in \cref{fig:comp} above, which shows that the new constraints derived in this paper complement the existing ones from direct searches for dark matter--nucleus scattering in underground detectors and cover significant unexplored parameter space.  Our results are also complementary to prior work utilizing the Earth's heat flow to set limits on annihilating dark matter~\cite{Mack:2007xj, Bramante:2019fhi}

In the future, it would be interesting to consider similar signatures of non-annihilating dark matter in exoplanets \cite{Leane:2020wob} and in stars other than the Sun. In particular, observations of Red Giants and Brown Dwarfs might enlarge the bounds we have obtained here, although they will depend in some detail on the internal isotopic composition of these stars. We leave this and other applications of benign and malignant black holes formed from dark matter to future work.

\section*{Acknowledgements}

We thank Amit Bhoonah, Jonathan Corbett, Simon Knapen, Ranjan Laha, Nirmal Raj, Ningqiang Song, Aaron Vincent, and Tevong You for useful conversations. The work of JA, JB, AG is supported by the Natural Sciences and Engineering Research Council of Canada (NSERC).  J.K.\ has been partially supported by the European Research Council (ERC) under the European Union's Horizon 2020 research and innovation program (grant agreement No.\ 637506, ``$\nu$Directions'') and by the German Research Foundation (DFG) under grant No.\ KO~4820/4-1. Research at Perimeter Institute is supported in part by the Government of Canada through the Department of Innovation, Science and Economic Development Canada and by the Province of Ontario through the Ministry of Colleges and Universities.

\appendix
\section{Capture rate scaling}
\label{app:cap}

In this appendix, we give more details on the way in which our estimate of the dark matter capture rate differs from past work in Ref.~\cite{Albuquerque:2000rk}. The main addition we have implemented in \cref{eq:boltzmannShift} compared to Ref.~\cite{Albuquerque:2000rk} is the boost that dark matter particles obtain as they fall into the gravitational potential of the Sun or Earth.

To understand the impact of this correction, let us contrast the uncorrected flux-normalized dark matter velocity distribution, $f(v)$, to the corrected one, $f(\sqrt{v^2 - v_e^2}) \equiv f(u)$. For clarity, in this appendix we treat these distributions as function of $v = |\vec{v}|$ only, neglecting the angular dependence arising from the solar system's motion relative to the Milky Way's dark matter halo.  When $\sigma_{\chi N}$ is large so that an $\mathcal{O}(1)$ fraction of all dark matter particles crossing the Sun or Earth is being captured, the number of captured dark matter particles is nearly independent of the gravitational boost, that is $\int_{v_e}^{v_\text{max}} \! d^3v \, f(u) \simeq \int_{v_e}^{v_\text{max}} \! d^3v \, f(v)$. The reason is that in this case, $v \gg v_e$ over most of the integration region, and thus $\sqrt{v^2-v_e^2} \simeq v$.  However, when $\sigma_{\chi N}$ is small, and therefore the only the slowest dark matter particles can be captured, we see a significant difference. This can be understood by considering the limit $u \to 0$ (and therefore $v \to v_e$), where we find that
\begin{align}
    f(u) |_{v \to v_e} \sim v_e^{3/2} u^{3/2} \,, \qquad\qquad f(v) |_{v \to v_e} \sim v^{3}.
    \label{eq:fExpand}
\end{align}
We can see the results of this difference when comparing our capture rates plotted in \cref{fig:CaptureRatePlots} to Fig.~1 of Albuquerque et al.~\cite{Albuquerque:2000rk}. The number of dark matter particles captured per second ($C_\chi$) is the same between these two results at high $\sigma_{\chi N}$, but differs at lower $\sigma_{\chi N}$ as slopes converge to their asymptotic values from \cref{eq:fExpand}.

As for the exclusion limits on $\sigma_{\chi N}$ as a function of $m_\chi$, \cref{eq:vmax,eq:fExpand} together tell us that
\begin{align}
    u_\text{max}^{(v\to v_e)} \propto \frac{\sigma_{\chi N}}{m_\chi} \,,
    \label{eq:vmaxTaylor}
\end{align}
where we work in the limit $m_\chi \gg m_j$ and we use the notation from \cref{sec:capture-rate}, with $L$ the distance traveled inside the Sun or Earth, $m_j$ the mass of the target nuclei, and $n_j$ their number density. Integrating over the dark matter velocity distribution to obtain the capture probability as in \cref{eq:Pcap,eq:Mcap}, but using our new definitions, we find that if we use $f(u)$, the captured dark matter mass after a given time interval obeys the scaling
\begin{align}
    M_\text{cap}^{(f(u))} &\propto \bigg( \frac{\sigma_{\chi N}}{m_\chi} \bigg)^{5/2} \,,
    \label{eq:McapTayloru}
\intertext{and if we use $f(v)$, it scales as}
    M_\text{cap}^{(f(v))} &\propto \bigg( \frac{\sigma_{\chi N}}{m_\chi}\bigg)^{4}.
    \label{eq:McapTaylorv}
\end{align}
The scaling behavior of $M_\text{cap}^{(f(u))}$ is consistent with the scaling observed in \cref{fig:CaptureRatePlots} and the corresponding fitting functions, \cref{eq:C-fit-1,eq:C-fit-2,eq:C-fit-3,eq:C-fit-4}, in the high-mass limit once we take into account the fact that the captured mass is proportional to the capture rate times the DM masse, $M_\text{cap} \propto C_\chi m_\chi$.

Recalling from \cref{sec:collapse-conditions} that the critical mass for collapse scales as $M_\text{crit} \propto m_\chi^{-3/2}$ (in the absence of Pauli blocking), and equating $M_\text{cap}$ to $M_\text{crit}$, we can find the minimum excluded cross-section at large $m_\chi$ (and therefore low capture probability). It turns out to scale as
\begin{align}
    \sigma_{\chi N}^{(f(u))} \propto m_\chi^{2/5}
    \label{eq:sigmaU}
\intertext{when using $f(u)$ and thus taking into account dark matter acceleration in the Sun's or Earth's gravitational field, and as}
    \sigma_{\chi N}^{(f(v))} \sim m_\chi^{5/8}
    \label{eq:sigmaV}
\end{align}
when this extra acceleration is neglected and we simply work with the uncorrected $f(v)$.

\section{Velocity scaling of the energy loss rate}
\label{app:collapse}

In this appendix, we justify how the velocity scaling of the energy loss rate in the inertial and viscous regime is obtained, cf.\ \cref{eq:dEdt-inertial-1,eq:dEdt-viscous-1}, respectively. For simplicity, we consider one-dimensional motion and non-relativistic kinematics. The full three-dimensional treatment using a Maxwell velocity distribution for the nuclei yields the same velocity scalings in the energy transfer rate, modulo an $\sim$ $\mathcal{O}(1)$ factor from the angular integration. 

\begin{figure}
    \centering
    \includegraphics[width=0.8\linewidth]{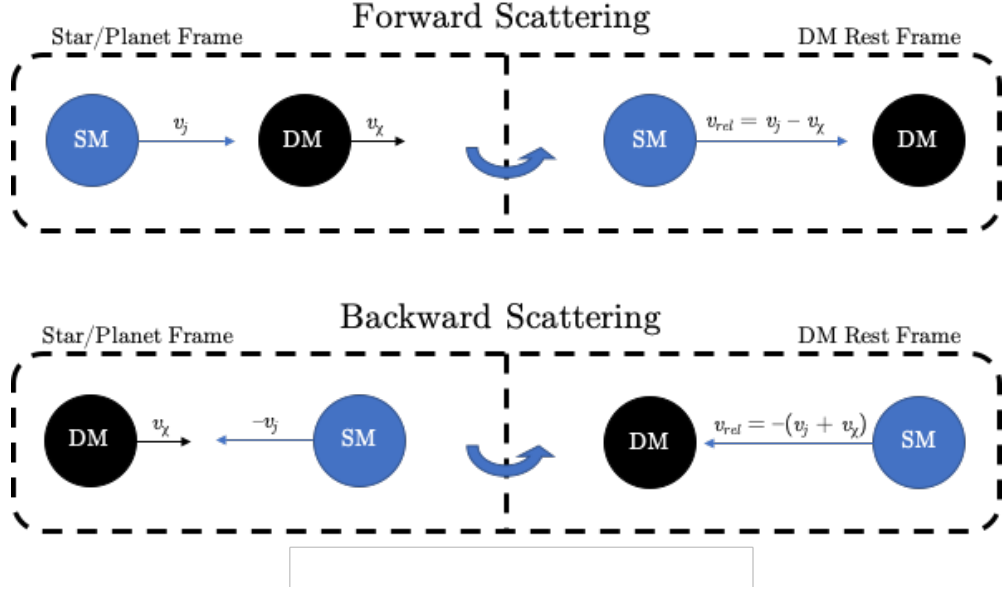}
    \caption{A schematic describing the one-dimensional kinematics of DM--nucleus scattering used in this appendix, which details the velocity scalings for DM energy loss in viscous and inertial regimes.  All diagrams depict the initial state of the scattering process, both in the stellar or planetary rest frame (left) and in the dark matter particle's rest frame (right).}
    \label{fig:Scattering}
\end{figure}

It is convenient to introduce two different frames: the star or planetary rest frame, and the dark matter rest frame. Depending on the relative motion of the dark matter particle with respect to nuclei, it can either forward-scatter (increasing its velocity) or back-scatter (decreasing or reversing its velocity), as illustrated in \cref{fig:Scattering}. The energy transfer in each process is computed assuming the scattering is elastic. Forward scattering occurs only when the dark matter particle is moving slower than the thermal motion of the nuclei. As we show below, this leads to the velocity scaling $\sim v_{\chi}^2 v_j$ in the viscous regime, cf.\ \cref{eq:dEdt-viscous-1}. 

In the dark matter rest frame, conservation of the total kinetic energy dictates
\begin{align}
    \frac{p_j^2}{2m_j}=\frac{p_j'^2}{2m_j} + \frac{p_\chi'^2}{2m_\chi},,
    \label{eq:consEnergy}
\end{align}
\noindent where $m_j$, $p_j = m_j v_j$, and $p_j' = m_j v_j'$ are, respectively, the mass, initial momentum, and final momentum of the target nucleus, $m_\chi$ is the dark matter mass, and $p_\chi' = m_\chi v_\chi'$ is the dark matter's final momentum. On the other hand, conservation of momentum imposes
\begin{align}
    p_j = p_j' + p_\chi' \,.
    \label{eq:consMomentum}
\end{align}
Combining \cref{eq:consEnergy,eq:consMomentum} yields
\begin{align}
    \frac{p_j^2}{2m_j}=\frac{(p_j-p_\chi')^2}{2m_j} + \frac{p_\chi'^2}{2m_\chi},
    \label{eq:fullCons}
\end{align}
from which we solve for the dark matter's final velocity,
\begin{align}
    2 v_j = m_\chi v_\chi' \left(\frac{1}{m_j} + \frac{1}{m_\chi} \right)
        \quad\to\quad  v_\chi' = \frac{2 v_j}{m_\chi}\mu(m_j) \,.
    \label{eq:vx}
\end{align}
Here, the quantity $\mu(m_j)$ is the reduced mass of the dark matter--nucleus system.

Boosting \cref{eq:vx} back to the stellar or planetary rest frame gives
\begin{align}
    v_\chi' = v_\chi + \frac{2 (v_j - v_\chi)}{m_\chi}\mu(m_j) ,,
    \label{eq:vxprime}
\end{align}
and from this expression, we can compute the energy transfer in a single scatter in the stellar or planetary rest frame:
\begin{align}
    \Delta E = \frac{m_\chi}{2} (v_\chi'^2 - v_\chi^2)
      &= \frac{m_\chi}{2} \bigg( \frac{4 v_\chi(v_j - v_\chi)}{m_\chi} \mu(m_j)
                               + \frac{4 (v_j - v_\chi)^2}{m_\chi^2} [\mu(m_j)]^2 \bigg)
                                     \notag\\
      &= 2(v_j-v_\chi) \bigg( v_\chi + \frac{v_j-v_\chi}{m_\chi}\mu(m_j) \bigg) \mu(m_j) \,.
\end{align}
The energy loss rate is approximately given by the product of the scattering frequency $n_j \sigma_{\chi j} |v_\text{rel}|$ multiplied by the energy loss per scatter, i.e.\ $\dot{E} = n_j\sigma_{\chi j} |v_\text{rel}|\Delta E$. Here, $|v_\text{rel}| = |v_j - v_\chi|$ is the relative velocity in the DM--nucleus system.  Since the dark matter enters the viscous regime in the final stages of thermalization, $n_j$ and $T$ are the number density and temperature at the core of the stellar object. We obtain
\begin{align}
    \frac{dE}{dt} = 2 n_j\sigma_{\chi j} |v_j-v_\chi| (v_j-v_\chi) \left( v_\chi + \frac{v_j-v_\chi}{m_\chi}\mu(m_j) \right)\mu(m_j) \,.
    \label{eq:dEdt-full}
\end{align}
In the case that $v_j \ll v_\chi$ (inertial regime), we can drop $v_j$ from this expression. Assuming moreover $m_\chi \gg m_j$, this leads to
\begin{align}
    \frac{dE}{dt} \sim -m_j n_j \sigma_{\chi j} v_\chi^3 \,,
\end{align}
justifying \cref{eq:dEdt-inertial-1}.

In contrast, when the dark matter is moving slowly compared to the thermal speed of the nuclei, \cref{eq:dEdt-full} can be simplified by using $|v_j| \gg v_\chi$, but also noting that the average over all target nuclei, $\ev{v_j}$, vanishes. As we are really interested in the average energy loss rate, we can therefore drop all terms that change sign under the replacement $v_j \to -v_j$, and replace $v_j$ by its average thermal value $v_\text{th}^{(j)} = \sqrt{3 T / m_j}$ in all other terms. The result is
\begin{align}
    \ev{\frac{dE}{dt}} \sim -m_j n_j \sigma_{\chi j} v_\text{th}^{(j)} v_\chi^2 \,,
\end{align}
reproducing \cref{eq:dEdt-viscous-1}. In \cref{eq:dEdt-viscous-1}, we have not explicitly indicated the thermal average $\ev{\cdot}$ to simplify the notation.

\bibliographystyle{JHEP.bst}

\let\oldaddcontentsline\addcontentsline
\renewcommand{\addcontentsline}[3]{}

\bibliography{mybibliography}

\let\addcontentsline\oldaddcontentsline
\end{document}